\documentclass[prb,twocolumn,showpacs]{revtex4}
\usepackage{color,soul}
\usepackage{amsmath}
\usepackage{dcolumn}
\usepackage{graphicx}
\usepackage{bm}
\usepackage{amssymb}
\usepackage{subfigure}
\usepackage{diagbox}
\begin{document}

\title{Closely competing valence bond crystal orders in the ground state of the spin-$\frac{1}{2}$ antiferromagnetic Heisenberg model on the pyrochlore lattice: a large scale unrestricted variational study}
\author{Rong Cheng and Tao Li}
\email{litao_phys@ruc.edu.cn}
\affiliation{School of Physics, Renmin University of China, Beijing 100872, P.R.China}

\begin{abstract}
The spin-$\frac{1}{2}$ antiferromagnetic Heisenberg model on the pyrochlore lattice(PAFH) is arguably the most well known strongly frustrated quantum magnet in three spatial dimension. However, due to the rapid scaling of Hilbert space with the linear size of such a three dimensional system, most previous studies of this model are conducted on rather small clusters and the nature of its ground state in the thermodynamic limit remains elusive after about 30 years' intense debate. Here we apply a recently developed powerful algorithm to perform large scale unrestricted variational optimization of the ground state of the spin-$\frac{1}{2}$ PAFH from the perspective of the resonating valence bond(RVB) theory. The largest cluster that we have attempted contains as many as 2048 sites and the corresponding variational wave function contains as many as 16777216 variational parameters. These are both much larger than that had ever been attempted. Through such a large scale variational optimization, we find a highly competitive candidate ground state for the spin-$\frac{1}{2}$ PAFH. This novel state features a maximally resonating valence bond crystal(VBC) pattern with $2\vec{a}_{1}\times2\vec{a}_{2}\times2\vec{a}_{3}$ periodicity. There are at least four levels of hierarchical structure in such a VBC state, with the first and the second level of hierarchy related to the breaking of the inversion and the translational symmetry. As a result, there is strong disparity in the extent of ice rule violation between up and down tetrahedrons in this state. Intriguingly, we find that within the RVB framework such a maximally resonating VBC state is almost degenerate with a recently proposed VBC state obtained from dressing hard hexagon covering of the pyrochlore lattice, although the extent of ice rule violation is totally uniform in the latter case. We find that further symmetry breaking will occur in the dressed hard hexagon VBC state under unrestricted optimization. In particular, strong disparity in the extent of ice rule violation for up and down tetrahedrons will also emerge in this state. We show that the maximally resonating VBC state found here will be favored by a tiny next-neighboring exchange coupling over the dressed hard hexagon VBC state.             
\end{abstract}

\maketitle

\section{Introduction}
Strongly frustrated quantum magnet is a marvelous playground to explore novel state of matter and novel form of excitations. Study of these issues are not only for the fun of fundamental research, but may also have great potential in future applications, especially that related to quantum information industry. The spin-$\frac{1}{2}$ antiferromagnetic Heisenberg model defined on the kagome lattice(KAFH) is probably the most well known example of strongly frustrated quantum magnet in two spatial dimension and it has been the focus of intensive research in the last three decades both theoretically and experimentally. While there remains a large amount of mysteries with this system, it is generally believed that the ground state of the spin-$\frac{1}{2}$ KAFH host a quantum spin liquid\cite{Elser,Chalker,Leung,Sindzingre,Nakano,Lauchli,Jiang,Yan,Hastings,Ran,Iqbal,Liao}.

The spin-$\frac{1}{2}$ antiferromagnetic Heisenberg model defined on the pyrochlore lattice(PAFH) is a close analogy of the spin-$\frac{1}{2}$ KAFH in three spatial dimension and it has long been the focus of theoretical studies\cite{Harris,Isoda,Canals1,Canals2,Koga,Tsunetsugu1,Tsunetsugu2,Berg,Sondhi1,Sondhi2,Han,Burnell,Normand1,Normand2,Huang,Chandra,Iqbal2,Liu,Luitz,Nikita,Robin}. The pyrochlore lattice is a three dimensional network made of conner-sharing tetrahedrons. It is closely related to the  kagome lattice, which is a two dimensional network made of conner-sharing triangles. More specifically, the pyrochlore lattice can be viewed as the stacking of kagome planes oriented in the direction of the four facets of an elementary tetrahedron in the pyrochlore lattice(see Fig.1). The Hamiltonian of the spin-$\frac{1}{2}$ PAFH is given by
\begin{equation}
H_{J}=J\sum_{<i,j>}\hat{\mathbf{s}}_{i}\cdot\hat{\mathbf{s}}_{j}
\end{equation}
Here the sum is over nearest-neighboring(NN) bonds of the pyrochlore lattice. $J>0$ is the antiferromagnetic exchange coupling between NN sites. In the following, we will set $J=1$ as the unit of energy. The strongly frustrated nature of the spin-$\frac{1}{2}$ KAFH and the spin-$\frac{1}{2}$ PAFH can be seen more clearly by rewriting their Hamiltonian as the sum of squares of the total spin in their structure units 
\begin{equation}
\hat{H}=\frac{J}{2}\sum_{u}\hat{\mathbf{S}}_{u}^{2}+C
\end{equation}
in which $u$ denotes the structure unit in the corresponding lattice, namely triangles in the kagome lattice and tetrahedrons in the pyrochlore lattice. 
\begin{equation}
\hat{\mathbf{S}}_{u}=\sum_{i\in u}\hat{\mathbf{s}}_{i}
\end{equation}
denotes the total spin in structure unit $u$, $C$ is an unimportant constant. We note that the dual lattice(the lattice made up of the geometric center of the structure units) of the kagome lattice and the pyrochlore lattice, namely the honeycomb lattice and the diamond lattice, are both bipartite. Thus the structure unit on both lattices fall into two categories. In the following, we will call these two types of structure unit the "up" and "down" triangle(tetrahedron). 

\begin{figure}
\includegraphics[width=8.5cm]{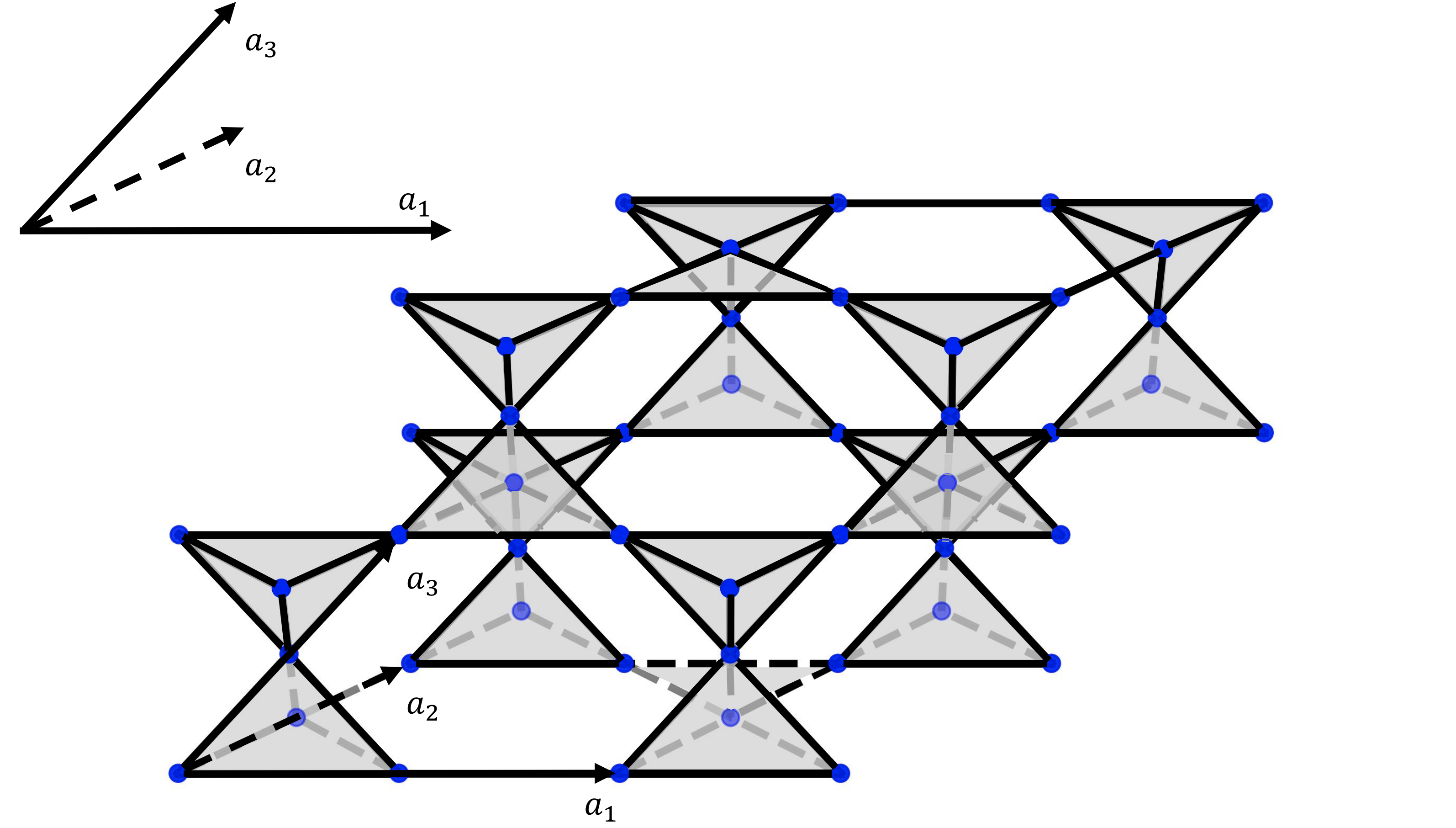}
\caption{Illustration of the pyrochlore lattice. The pyrochlore lattice is a three dimensional network made of conner-sharing tetrahedrons. $a_{1},a_{2},a_{3}$ denote the three basis vectors of the pyrochlore lattice. The pyrochlore lattice can also be viewed as the stacking of kagome planes oriented in the direction of the four facets of an elementary tetrahedron, namely in the $[1,0,0]$, $[0,1,0]$, $[0,0,1]$ and the $[1,1,1]$ direction. }
\end{figure}

Obviously, if $\hat{\mathbf{S}}_{u}^{2}$ can be minimized simultaneously on all structure units in a quantum state, then this state must be the ground state of the system. 
In the semiclassical limit of $S\rightarrow\infty$, this condition can indeed be fulfilled on the kagome and the pyrochlore lattice\cite{Chalker1,Chalker2,Chalker3}. More specifically, there are sub-extensive(kagome) or extensive(pyrochlore) semiclassical spin configurations that can satisfy the following ice rule condition
\begin{equation}
\mathbf{S}_{u}\equiv 0
\end{equation} 
The existence of such sub-extensive or extensive degeneracy in the classical ground state implies that there is no preferred semiclassical ordering pattern in these strongly frustrated magnets. Indeed, classical spin liquid behavior(or collective paramagnetic behavior) has been intensively studied in the classical limit of these models over the last three decades\cite{Chalker1,Chalker2,Chalker3,ShiAC}.The strong frustration effect as argued above from the semiclassical perspective place the spin-$\frac{1}{2}$ KAFH and the spin-$\frac{1}{2}$ PAFH at a favorable position to realize a quantum spin liquid ground state.

The situation in the extreme quantum limit with a $S=\frac{1}{2}$ spin is however subtly different. While the quantum counterpart of the ice rule, namely
\begin{equation}
\hat{\mathbf{S}}_{u}=0
\end{equation} 
can be fulfilled locally in a single tetrahedron of the pyrochlore lattice, this can never happen simultaneously on all tetrahedrons of the lattice. More specifically, the ice rule can be satisfied simultaneously on at most one half of all tetrahedrons, namely either on all the up-tetrahedrons or on all the down-tetrahedrons. The situation of the spin-$\frac{1}{2}$ KAFH is even more subtle since the condition of $\hat{\mathbf{S}}_{u}=0$ can never be satisfied on any triangle. In fact, the minimal value of $\hat{\mathbf{S}}^{2}_{u}$ on a triangle is $\frac{3}{4}$. At the same time, only $\frac{3}{4}$ of all triangles of the kagome lattice can reach this minimal value simultaneously\cite{Elser}. Such a subtle difference between the spin-$\frac{1}{2}$ KAFH and the spin-$\frac{1}{2}$ PAFH may lead to ground state of very different nature.

As for the spin-$\frac{1}{2}$ KAFH, it is generally believed that a quantum spin liquid is realized in its ground state. This belief is supported by a large amount of evidences from both the theoretical and the experimental side. On the theoretical side, while there remains controversy on some details, calculation adopting a broad spectrum of methods ranging from exact diagonalization\cite{Elser,Chalker,Leung,Sindzingre,Nakano,Lauchli}, variational Monte Carlo\cite{Hastings,Ran,Iqbal}, DMRG\cite{Jiang,Yan} to Tensor RG\cite{Liao} all agree on the spin liquid nature of the ground state. On the experimental side, inelastic neutron scattering\cite{Helton,HanTH}, muon spin rotation\cite{Mendels} and nuclear magnetic resonance measurement\cite{Mendels1,Fu,Khunita} on kagome spin liquid candidate materials, for example, the Herbertsmithite ZnCu$_{3}$(OH)$_{6}$Cl$_{2}$\cite{Mendels,Helton,HanTH}, the Zn-Barlowite ZnCu$_{3}$(OH)$_{6}$FrBr\cite{Shi}, and the more recently synthesized YCu$_{3}$(OH)$_{6}$Br$_{2}$[Br$_{1-x}$(OH)$_{x}$](YCOB)\cite{XHChen} are all consistent with the existence of a quantum spin liquid ground state, although the exact nature of the spin liquid is again under debate. On the other hand, the case for a quantum spin liquid ground state in the spin-$\frac{1}{2}$ PAFH is much less clear, even though the study of both models have lasted for three decades. Theoretically, suggestions about the ground state ranging from a novel quantum spin liquid state to various kinds of symmetry broken valence bond crystal(VBC) state have been proposed over the years\cite{Harris,Isoda,Canals1,Canals2,Koga,Tsunetsugu1,Tsunetsugu2,Berg,Sondhi1,Sondhi2,Han,Burnell,Normand1,Normand2,Huang,Chandra,Iqbal2,Liu,Luitz,Nikita}. On the experimental side, the study of pyrochlore antiferromagnet has long been dominated by the study of its anisotropic version\cite{Moessner,Ross,Hermele,Shannon,HYan}, in particular the study of the classical and quantum spin ice behavior\cite{spinice1,spinice2,spinice3} in the 227 material Dy$_{2}$Ti$_{2}$O$_{7}$ and Ho$_{2}$Ti$_{2}$O$_{7}$. It is until recently that the Heisenberg limit of the model has been realized in the rare earth molybdenum oxynitride Lu$_{2}$Mo$_{2}$O$_{5}$N$_{2}$, in which the spin-$\frac{1}{2}$ Mo$^{5+}$ ions occupying the pyrochlore lattice site is believed to have a dominant antiferromagnetic Heisenberg exchange coupling between nearest-neighboring sites\cite{Clark,Iqbal3}. Measurements on this material find that the system remains paramagnetic at a temperature scale much smaller than that set by the NN exchange coupling, implying the possibility of a quantum spin liquid ground state.

The short of consensus on the nature of the ground state of the spin-$\frac{1}{2}$ PAFH can be mainly attributed to the three dimensionality of the pyrochlore lattice. A direct consequence of this is the rapid scaling of the Hilbert space with the linear system size. As a result, most theoretical calculations on the spin-$\frac{1}{2}$ PAFH are done on rather small clusters. For example, exact diagonalization result on the spin-$\frac{1}{2}$ PAFH is only available on cluster as small as $N=2^{3}\times4$\cite{Canals1,Canals2}. The result obtained on such small cluster can be rather misleading when we want to extract the information about the ground state of the system in the thermodynamic limit. In fact, as we will show in this paper, the ground state of the spin-$\frac{1}{2}$ PAFH on the $2^{3}\times4$ cluster is qualitatively different from that on larger clusters. Again, as a result of the three dimensionality of the pyrochlore lattice, the DMRG method that is extensively used in the study of one and two dimensional quantum magnets, runs out of potential due to the ultra-fast scaling of entanglement entropy with the linear system size. In fact, the largest DMRG calculation that have been attempted on the spin-$\frac{1}{2}$ PAFH is done on a $N=3^{3}\times4$ cluster\cite{Luitz}. For the same reason, the tensor network RG method may also have a tough time dealing with such a three dimensional system. Compared to these methods, the variational Monte Carlo(VMC) method is less restrictive on system size. However, even the VMC study of the spin-$\frac{1}{2}$ PAFH has only been attempted on rather small clusters\cite{Han,Burnell,Nikita}. For example, in a recent VMC study of the spin-$\frac{1}{2}$ PAFH\cite{Nikita}, the largest pyrochlore cluster used has a size of only  $N=4^{3}\times 4$. 
  
Beside the rapid scaling of the Hilbert space with the linear system size, an additional challenge in the study of the spin-$\frac{1}{2}$ PAFH is the intricate competition between different potential symmetry breaking patterns in the ground state of such a strongly frustrated system. As a result, we usually can not adopt a simple ansatz for the ground state with presumed symmetry. This is recently shown to be crucial in the study of two typical strongly frustrated quantum magnets. In Ref.[\onlinecite{Tao1}], we find that the well known $U(1)$ spin liquid with a large spinon fermi surface(SFS) is not the ground state of the spin-$\frac{1}{2}$ antiferromagnetic Heisenberg model on the triangular lattice supplemented with strong four spin ring exchange coupling(the J$_{1}$-J$_{4}$ model). Instead, we find that the true ground state of the model features a complicated VBC pattern with $4\times6$ periodicity in the parameter regime usually expected for the $U(1)$ SFS state. In Ref.[\onlinecite{Tao2}], we find that the $1/9$ magnetization plateau of the spin-$\frac{1}{2}$ KAFH features a VBC pattern with $3\times3$ periodicity and an enlarged unit cell containing $27$ sites. These discoveries are made possible by a new optimization algorithm designed for large scale unrestricted optimization of resonating valence bond(RVB) wave function for strongly frustrated quantum magnet. The essence of the new algorithm is to approximate the Hessen matrix of the problem with a finite-depth BFGS iteration. Such a finite-depth BFGS algorithm can achieve good balance between numerical efficiency, numerical stability and storage demand and make possible the optimization of millions of variational parameters with relatively small amount of computational resources.  

In this work, we apply the finite-depth BFGS algorithm to investigate the ground state of the spin-$\frac{1}{2}$ PAFH. For this purpose, we propose the most general RVB ansatz that is consistent with the spin symmetry of the system. No further assumption is made on such a generalized RVB ansatz, which contains $4N^{2}$ variational parameters on a finite cluster with $N$ site. Through unrestricted variational optimization on such a generalized RVB ansatz, we find a candidate ground state of the spin-$\frac{1}{2}$ PAFH that features an intricate VBC pattern with $2\vec{a}_{1}\times2\vec{a}_{2}\times2\vec{a}_{3}$ periodicity. We find that there are at least four levels of hierarchical structure in such a VBC pattern. At the first level of the hierarchy, we find that all the down(or equivalently all the up) tetrahedrons are strongly spin dimerized. The inversion symmetry relating the up and the down tetrahedron is thus broken. Accompanying this is the significant reduction of $\langle\hat{\mathbf{S}}_{u}^{2}\rangle$ in the down tetrahedron as compared to that in the up tetrahedron. At the second level of the hierarchy, $\frac{3}{4}$ of down tetrahedrons are involved in $6$-spin resonance process along hexagonal rings. Such resonating rings are made up of strong valence bonds in down tetrahedrons and moderate valence bonds in up tetrahedrons. As the VBC pattern we find maximizes the number of such resonating hexagonal ring, we called it the maximally resonating VBC pattern. The translational symmetry is broken as a result of the maximally resonating pattern and the unit cell is doubled along all the three basis vector directions of the pyrochlore lattice. Above these two base levels, we find that there are still higher level of hierarchical structures in the maximally resonating VBC pattern. While these higher order structure contribute little to the ground state energy, they are clearly resolved from our unrestricted optimization of the generalized RVB ansatz. 

We find that an NN-RVB ansatz with $2\times 2\times 2$ periodicity can capture very well the qualitative feature of the maximally resonating VBC state. With the help of the NN-RVB ansatz, we can extend our calculation to very large clusters. The ground state energy obtained from the NN-RVB ansatz and the generalized RVB ansatz are very close to each other. They extrapolate to $-0.4827J/site$ and $-0.4846J/site$ respectively in the thermodynamic limit. These results are obtained from calculation on clusters containing as many as $N=8^{3}\times4=2048$ site and with wave function containing as many as $N_{v}=16777216$ variational parameters. This is much larger than any previous attempt on the same model. As a byproduct, we have also falsified the monopole flux phase proposed in Ref.[\onlinecite{Han}] and Ref.[\onlinecite{Burnell}], whose variational energy is found to extrapolate to $-0.4574J/site$ in the thermodynamic limit. 
 
Very recently, a VBC state obtained from dressing hard hexagon covering of the pyrochlore lattice is found to have a very good variational energy\cite{Robin,Cheong}. Through series expansion on the dressing parameter, the energy of such a VBC state is found to be $-0.48947J/site$ in the thermodynamic limit. This sets the lowest known upper bound on the ground state energy of the spin-$\frac{1}{2}$ PAFH. The structure of this VBC state is drastically different from the maximally resonating VBC state found in this work. More specifically, while the maximally resonating VBC state has a unit cell containing 32 sites, the unit cell of the dressed hard hexagon VBC state contains at least $48$ sites. At the same time, $\langle\hat{\mathbf{S}}_{u}^{2}\rangle$ remains uniform on all tetrahedrons in the dressed hard hexagon VBC state, although inversion, rotation and translation symmetry are indeed all broken. In fact, the dressed hard hexagon VBC state proposed in Ref.[\onlinecite{Robin}] has the same symmetry as the bare hard hexagon covering of the pyrochlore lattice, which is invariant under the combined operation of spatial inversion and translation or rotation. 

To decide the relative stability of the maximally resonating VBC state and the dressed hard hexagon VBC state, we must compute their energy at the same approximation level and under the same condition. For this purpose, we have performed variational optimization on pyrochlore cluster large enough to accommodate both types of VBC state. We find the smallest equilateral pyrochore cluster(pyrochore cluster with the geometry of $4\times L^{3}$) that can meet such a requirement contains $N=4\times 6^{3}=864$ sites. Our refined variational optimization on the $L=6$ cluster indicates that while both VBC states have very different structure, they are essentially degenerate. Although we are still lack of a simple interpretation for such an astonishing degeneracy, we note both types of VBC state feature 6-spin resonance on hexagonal rings. The key difference between the two states is as follows. While all sites are involved in such resonating rings and each site participates in only one resonating ring in the dressed hard hexagon VBC state, in our maximally resonating VBC state a site can participate in more than one resonating rings simultaneously, although not all sites are involved such rings. In addition, we find that further symmetry breaking will occur in the dressed hard hexagon VBC state with unrestricted optimization. In particular, the value of $\langle\hat{\mathbf{S}}_{u}^{2}\rangle$ becomes different for up and down tetrahedrons in the fully optimized state. Thus, the invariance of the hard hexagon covering under the combined operation of spatial inversion and translation or rotation will eventually be lost. Finally, we note that the maximally resonating VBC state will be favored over the dressed hard hexagon VBC state when a tiny next-neighboring exchange coupling is turned on.

The paper is organized as follows. In the next section, we introduce the two different forms of RVB wave function used to describe the ground state of the spin-$\frac{1}{2}$ PAFH in this work and the related variational optimization algorithm. In Sec.III, we present the numerical results of our variational calculation. The last section of the paper is devoted to the conclusion of our results and a discussion on the relevance of these results to the experimental observations made on recently synthesized pyrochlore quantum spin liquid candidate materials.

\section{The RVB wave functions for the ground state of the spin-$\frac{1}{2}$ PAFH and their optimization}

In this work, we describe the ground state of the spin-$\frac{1}{2}$ PAFH within the fermionic RVB theory scheme. The variational wave function for the ground state is constructed by introducing slave particle representation of the spin operator as follows
\begin{equation}
\mathbf{S}_{i}=\frac{1}{2}\sum_{\alpha,\beta=\uparrow,\downarrow}f^{\dagger}_{i,\alpha}\bm{\sigma}_{\alpha,\beta}f_{i,\beta}
\end{equation}
Here $f_{i,\alpha}$ denotes the annihilation operator of a fermionic spinon with spin $\alpha$ at site $i$. $\bm{\sigma}_{\alpha,\beta}$ is the matrix element of the usual spin Pauli matrix. To represent faithfully the spin algebra of a $S=1/2$ quantum spin, the slave particle should be subjected to the following no double occupancy constraint
\begin{equation} 
\sum_{\alpha=\uparrow,\downarrow}f^{\dagger}_{i,\alpha}f_{i,\alpha}=1
\end{equation}
The hamiltonian of the spin-$\frac{1}{2}$ PAFH written in terms of the spinon operator reads
\begin{eqnarray}
H&=&\frac{J}{4}\sum_{\langle i,j \rangle,\alpha,\beta,\gamma,\delta}f^{\dagger}_{i,\alpha}\bm{\sigma}_{\alpha,\beta}f_{i,\beta}\cdot f^{\dagger}_{j,\gamma}\bm{\sigma}_{\gamma,\delta}f_{j,\delta}\nonumber\\
&=&\frac{J}{2}\sum_{\langle i,j \rangle,\alpha,\beta}[f^{\dagger}_{i,\alpha}f_{i,\beta} f^{\dagger}_{j,\beta}f_{j,\alpha}-\frac{1}{2}f^{\dagger}_{i,\alpha}f_{i,\alpha} f^{\dagger}_{j,\beta}f_{j,\beta}]\nonumber\\
\end{eqnarray}
Here we have used the following identity for Pauli matrix
\begin{equation}
\bm{\sigma}_{\alpha,\beta}\cdot \bm{\sigma}_{\gamma,\delta}=2\delta_{\alpha,\delta}\delta_{\beta,\gamma}-\delta_{\alpha,\beta}\delta_{\gamma,\delta}
\end{equation}
The quartic term in the Hamiltonian can then be decoupled by introducing the following RVB mean field order parameters 
\begin{eqnarray}
\chi_{i,j}&=&\frac{3J}{8}\sum_{\alpha}\langle f^{\dagger}_{i,\alpha}f_{j,\alpha}\rangle\nonumber\\
\Delta_{i,j}&=&\frac{3J}{8}\sum_{\alpha}\langle f_{i,\alpha}f_{j,\bar{\alpha}}\rangle\nonumber\\
\end{eqnarray}
Here we have assumed spin rotational symmetry for both $\chi_{i,j}$ and $\Delta_{i,j}$ and $\bar{\alpha}$ denotes the opposite of $\alpha$. Both $\chi_{i,j}$ and $\Delta_{i,j}$ are in general complex numbers. Treating the spinon operator $f_{i,\alpha}$ as free fermion operator and perform the Wick decomposition we are led to the following RVB mean field Hamiltonian 
\begin{eqnarray}
H_{\mathrm{MF}}&=&-\sum_{\langle i,j \rangle,\alpha}[\chi_{i,j} f^{\dagger}_{i,\alpha}f_{j,\alpha}+\Delta_{i,j} f^{\dagger}_{i,\alpha}f^{\dagger}_{j,\bar{\alpha}}]+h.c.\nonumber\\
&+&\sum_{i,\alpha}\mu_{i}f^{\dagger}_{i,\alpha}f_{i,\alpha}
\end{eqnarray}
Here the onsite term is introduced to enforce the no double occupancy constraint at the mean field level. Eq.11 is called an RVB mean field ansatz. The RVB state that is used to describe the ground state of the spin-$\frac{1}{2}$ PAFH is generated by Gutzwiller projection of the mean field ground state of $H_{\mathrm{MF}}$
\begin{equation}
|f-\mathrm{RVB}\rangle=P_{G}|\mathrm{MF}\rangle
\end{equation}
here $|\mathrm{MF}\rangle$ is the ground state of $H_{\mathrm{MF}}$, $P_{G}$ is the Gutzwiller projection operator that removes the doubly occupied configuration from $|\mathrm{MF}\rangle$. The set of mean field order parameter $\{\chi_{i,j},\Delta_{i,j},\mu_{i} \}$ will be taken as the variational parameters of the ground state. 

We note that the hopping parameter $\chi_{i,j}$ and the pairing parameter $\Delta_{i,j}$ in $H_{\mathrm{MF}}$ are now all restricted to the nearest-neighboring(NN) bonds. We will thus call such an ansatz an NN-RVB mean field ansatz. More generally, we can relax the restriction on the range of the hopping parameter and the pairing parameter. The RVB state constructed from $H_{\mathrm{MF}}$ with arbitrarily long-ranged $\chi_{i,j}$ and $\Delta_{i,j}$ will be called a generalized RVB state. It represents the most general fermionic RVB state that is compatible with the spin symmetry of the system. Obviously, the generalized RVB state is more accurate than the NN-RVB state. However, the number of variational parameter in a generalized RVB state is also much larger than that in an NN-RVB state. More specifically, the total number of variational parameters in a generalized RVB state is of order $N^{2}$ on a finite cluster with $N$ site. On the other hand, the number of variational parameters in an NN-RVB state is only of order $N$. In our study, we will use both the NN-RVB state and the generalized RVB state as the variational ansatz to describe the ground state of the spin-$\frac{1}{2}$ PAFH. The generalized RVB ansatz is mainly used to extract the accurate structure of the ground state on small cluster, for example the periodicity of the symmetry breaking pattern in the ground state. The huge number of variational parameter in the generalized RVB state is then crucial to reduce the variational bias in the calculation. We will then use the NN-RVB ansatz with presumed periodicity to approximate these ground state structures on larger clusters. The optimized NN-RVB ansatz can also be used as initial guess for unrestricted optimization of the generalized RVB ansatz on large clusters.

The mean field ground state of $H_{\mathrm{MF}}$ can be found by rewriting it as the following matrix form
\begin{equation}
H_{\mathrm{MF}}=\bm{\psi}^{\dagger}\mathbf{M}\bm{\psi}
\end{equation}
here we have performed the following particle-hole transformation on the down-spin fermion operator
\begin{equation}
f^{\dagger}_{i,\downarrow}\rightarrow \tilde{f}_{i,\downarrow}
\end{equation}
$\bm{\psi}$ is given by
\begin{equation}
\bm{\psi}^{\dagger}=(f^{\dagger}_{1,\uparrow},...,f^{\dagger}_{N,\uparrow},\tilde{f}^{\dagger}_{1,\downarrow},....,\tilde{f}^{\dagger}_{N,\downarrow})
\end{equation}
$\mathbf{M}$ is a $2N\times 2N$ Hermitian matrix of the form
\begin{equation}
\mathbf{M}=-\left(\begin{array}{cc}\chi_{i,j} & \Delta_{i,j} \\ \Delta^{*}_{i,j} & -\chi_{j,i} \end{array}\right)+\left(\begin{array}{cc}\mu_{i}\delta_{i,j} & 0 \\ 0 & -\mu_{i}\delta_{i,j} \end{array}\right)
\end{equation}
$H_{\mathrm{MF}}$ can be taken into the following diagonalized form under unitary transformation 
 \begin{equation}
H_{\mathrm{MF}}=\sum_{i=1}^{2N}\epsilon_{i}\gamma^{\dagger}_{i}\gamma_{i}
\end{equation}
here
\begin{equation}
\bm{\gamma}=(\gamma_{1},...,\gamma_{N},\gamma_{N+1},....,\gamma_{2N})^{T}
\end{equation}
is related to $\bm{\psi}$ by the following unitary transformation 
\begin{equation}
\bm{\psi}=\mathbf{U}\bm{\gamma}
\end{equation} 
in which $\mathbf{U}$ is a $2N\times2N$ unitary matrix, $\epsilon_{i}$ is the $i$-th eigenvalue of matrix $\mathbf{M}$. We note that as a result of the spin rotational symmetry, these eigenvalues always appear in $\pm$ pairs. Here we assume that the first $N$ eigenvalues are negative. 

The mean field ground of $H_{\mathrm{MF}}$ then reads
\begin{equation}
|\mathrm{MF}\rangle=\prod_{i=1}^{N}\gamma^{\dagger}_{i}|\tilde{0}\rangle
\end{equation}
here $|\tilde{0}\rangle$ is the vacuum state of $\bm{\psi}$ and satisfy
\begin{equation}
f_{i,\uparrow}|\tilde{0}\rangle=f^{\dagger}_{i,\downarrow}|\tilde{0}\rangle=0
\end{equation}
In terms of the vacuum state of the original spinon operator $f_{i,\alpha}$, we have  
\begin{equation}
|\tilde{0}\rangle=\prod_{i=1}^{N}f^{\dagger}_{i,\downarrow}|0\rangle
\end{equation}
here $|0\rangle$ satisfies
\begin{equation}
f_{i,\uparrow}|0\rangle=f_{i,\downarrow}|0\rangle=0
\end{equation}

To perform variational Monte Carlo simulation on the RVB state $|f-\mathrm{RVB}\rangle$, we must choose a working basis. Here we adopt the conventional Ising basis as given by
\begin{equation}
|R\rangle=\prod^{N/2}_{k=1}f^{\dagger}_{i_{k},\uparrow}\tilde{f}^{\dagger}_{i_{k},\downarrow}|\tilde{0}\rangle=\prod^{N/2}_{k=1}\mathbf{S}^{+}_{i_{k}}|\tilde{0}\rangle
\end{equation}
in which $\{i_{1},....,i_{N/2} \}$ denotes the location of the $N/2$ up spin in the Ising basis $|R\rangle$, $\mathbf{S}^{+}_{i_{k}}$ denotes the spin raising operator on site $i_{k}$. The expansion of the fermionic RVB state in this Ising basis reads
\begin{equation}
|f-\mathrm{RVB}\rangle=\sum_{R}\Psi(R)|R\rangle
\end{equation}
The wave function $\Psi(R)$ of the RVB state in such an Ising basis is given by
\begin{equation}
\Psi(R)=\mathrm{Det}[\bm{\Phi}]
\end{equation}
in which $\bm{\Phi}$ is a $N\times N$ submatrix of $\mathbf{U}$ and is given by
\begin{equation}
\bm{\Phi}=\left(\begin{array}{cccc}U_{i_{1},1} & . & . & u_{i_{1},N} \\. & . & . & . \\U_{i_{N/2},1} & . & . & U_{i_{N/2},N} \\U_{i_{1}+N,1} & . & . & U_{i_{1}+N,N} \\. & . & . & .\\U_{i_{N/2}+N,1} & . & . & U_{i_{N/2}+N,N}\end{array}\right)
\end{equation}
Here $U_{i,j}$ is the matrix element of the unitary matrix $\mathbf{U}$. 

Following standard derivation\cite{Tao1}, the variational energy of the fermionic RVB state and its gradient with respect to the variational parameters are given by 
\begin{equation}
E=\langle H \rangle_{\Psi}=\frac{\langle\Psi| H |\Psi\rangle}{\langle \Psi |\Psi \rangle}=\frac{\sum_{R}|\Psi(R)|^2 E_{loc}(R)}{\sum_{R}|\Psi(R)|^2}
\end{equation}
and
\begin{equation}
\nabla E= \langle \nabla \ln \Psi(R) E_{loc}(R) \rangle_{\Psi}-E\langle \nabla \ln \Psi(R) \rangle_{\Psi}
\end{equation}
Here $E_{loc}(R)$ denotes the local energy in the Ising basis $|R\rangle$. It is given by
\begin{equation}
E_{loc}(R)=\sum_{R'}\langle R |H| R' \rangle \frac{\Psi(R')}{\Psi(R)}
\end{equation}
The expectation value involved in Eq.28 and Eq.29 can be computed by Monte Carlo sampling on the distribution generated by $|\Psi(R)|^2$. The gradient $\nabla \ln \Psi(R)$ in Eq.29 can be calculated as follows
\begin{equation}
\nabla \ln \Psi =\nabla \ln \mathrm{Det}[\bm{\Phi}]=\mathrm{Tr}[\nabla \bm{\Phi}\bm{\Phi}^{-1}]
\end{equation} 
The calculation of $\nabla \bm{\Phi}$, or $\nabla \mathbf{U}$, can be done using the first order perturbation theory. Denoting the $i$-th column of $\mathbf{U}$ as $\phi_{i}$, we have 
\begin{equation}
\nabla \phi_{i}=\sum_{\epsilon_{j}\neq\epsilon_{i}}\frac{\langle\phi_{j}|\nabla H_{MF}|\phi_{i}\rangle}{\epsilon_{i}-\epsilon_{j}}\phi_{j}
\end{equation}  
Here $\nabla H_{MF}$ denotes the gradient of the mean field Hamiltonian with respect to the variational parameters, $\langle\phi_{j}|\nabla H_{MF}|\phi_{i}\rangle$ denotes its matrix element between the $i$-th and the $j$-th eigenvector of the mean field Hamiltonian.

While such a computational procedure is indeed feasible for the NN-RVB ansatz even on relatively large cluster, it becomes prohibitively costly for the generalized RVB ansatz even for system of moderate size. Since the generalized RVB ansatz contains order of $N^{2}$ variational parameters, both the computation and the storage of $\nabla \mathbf{U}$ becomes extremely expensive when $N\geq 100$. One way out is to optimize the matrix element of $\mathbf{U}$ directly, instead of optimizing the mean field Hamiltonian that generate them. The main advantage of using the matrix element of $\mathbf{U}$ as the variational parameter is that the computation of $\nabla \bm{\Phi}$ becomes now almost trivial. More specifically, since the matrix element of $ \bm{\Phi}$ is now directly the variational parameter themselves, the matrix element of $\nabla \bm{\Phi}$ is either $1$ or $0$. Thus there is no need to compute or store $\nabla \mathbf{U}$ any more.  An additional advantage of using the matrix element of $\mathbf{U}$ as the variational parameter is that the so constructed RVB state may break the spin rotational symmetry and describe magnetic ordered state.

$\mathbf{U}$ is in general a complex matrix. Thus there will be $4N^{2}$ real parameters to be optimized in the generalized RVB ansatz. This number grows rapidly with the linear size of the system, particularly so for a three dimensional system. For example, on a pyrochlore cluster with the $L\times L\times L\times 4$ geometry and linear size $L=4$ the number of variational parameter in the generalized RVB ansatz is already $N_{v}=262144$. When the linear size of the system increases to $L=8$, this number will grow to $N_{v}=16777216$. Efficient optimization of so large number of parameters in variational Monte Carlo simulation is challenging. 

Recently, a finite-depth BFGS algorithm is developed for large scale unrestricted variational optimization\cite{Tao1,Tao2}. The key of the new algorithm is to generate an iterative approximation for the inverse of the Hessian matrix from the gradient of variational energy. More specifically, the approximate inverse Hessian matrix is updated as follows
\begin{equation}
\mathbf{B}_{k+1}=\left(\mathbf{I}-\frac{\mathbf{s}_{k}\mathbf{y}^{T}_{k}}{\mathbf{y}^{T}_{k}\mathbf{s}_{k} } \right)\mathbf{B}_{k}\left(\mathbf{I}-\frac{\mathbf{y}_{k}\mathbf{s}^{T}_{k}}{\mathbf{y}^{T}_{k}\mathbf{s}_{k} } \right)+\frac{\mathbf{s}_{k}\mathbf{s}^{T}_{k}}{\mathbf{y}^{T}_{k}\mathbf{s}_{k} }
\end{equation}
here $k=0,1,2.....$ is the iteration step index,
\begin{eqnarray}
\mathbf{s}_{k}&=&\bm{\alpha}_{k+1}-\bm{\alpha}_{k}\nonumber\\
\mathbf{y}_{k}&=&\nabla E_{k+1}-\nabla E_{k}
\end{eqnarray}  
are the difference between successive variational parameters and the difference between successive energy gradients at iteration step $k$. Here $\bm{\alpha}_{k=0}$ is our initial guess for the variational parameters. $\nabla E_{k=0}$ is the energy gradient calculated at the starting point. The inverse Hessian matrix $\mathbf{B}_{k}$ is initially set to be the identity matrix. Using such an iterative approximation on the inverse of the Hessian matrix, the variational parameters are updated as follows
\begin{equation}
\bm{\alpha}_{k+1}=\bm{\alpha}_{k}+\delta\ \mathbf{B}_{k} \nabla E_{k}
\end{equation}  
in which $\delta$ is the step length. It is usually chosen by trial and error. In practice, we restart the BFGS iteration every $K$ step. Since all we need in the BFGS iteration is the product of the matrix $\mathbf{B}_{k}$ with the vector $\nabla E_{k}$ or $\mathbf{y}_{k}$ and the inner product between the vector $\mathbf{s}_{k}$ and $\mathbf{y}_{k}$, we only need to store the $2K$ vectors $\mathbf{s}_{k}$ and $\mathbf{y}_{k}$ in the $K$-depth BFGS algorithm. The needed matrix-vector and vector inner product can be computed recursively using these $2K$ vectors. More details about the finite-depth BFGS algorithm can be found in Ref.[\onlinecite{Tao1}] and Ref.[\onlinecite{Tao2}].

\section{The maximally resonating VBC state as a candidate ground state of the spin-1/2 PAFH}
The variational optimization performed in this work is done on equilateral pyrocholore cluster with the $L\times L\times L\times 4$ geometry. Here $L$ denotes the linear size of the pyrochlore cluster, or the number of pyrochlore unit cell in each of the three basis directions. Periodic boundary condition is imposed on all the three basis directions. Unrestricted optimization of the generalized RVB ansatz is performed on the $L=2,4,6$ and $8$ cluster. For the $L=2$ and $L=4$ cluster, we start the optimization from a random initial guess. For the $L=6$ and the $L=8$ cluster, we start the optimization from an optimized NN-RVB ansatz with $2\times 2 \times 2$ periodicity. Such NN-RVB ansatz are obtained by extending the corresponding optimized NN-RVB ansatz on the $L=4$ cluster. We find that all these calculations lead consistently to the conclusion: the ground state of the spin-$\frac{1}{2}$ PAFH features a VBC pattern with $2\vec{a}_{1}\times2\vec{a}_{2}\times2\vec{a}_{3}$ periodicity. In the following, we will denote the number of lattice site of the pyrochlore cluster as $N$ and the number of pyrochlore unit cell as $N_{c}$. For the equilateral pyrochlore cluster studied in this work, we have $N=4L^{3}$ and $N_{c}=L^{3}$. The number of variational parameter in the variational ansatz will be denoted as $N_{v}$. We note that among the four equilateral pyrochlore clusters we have studied, only the $L=6$ cluster can accommodate the hard hexagon VBC state. Thus, a comparative study of the maximally resonating VBC state and the dressed hard hexagon VBC state will be carried out only on the $L=6$ cluster. 

\subsection{The ground state energy as obtained from unrestricted optimization of the generalized RVB ansatz}

We have performed unrestricted variational optimization of the generalized RVB ansatz for the spin-$\frac{1}{2}$ PAFH from random initial guess on both the $L=2$ and the $L=4$ pyrochlore cluster. The obtained ground state energy is $E_{0}\approx-0.5118J/site$ for the $L=2$ cluster and $E_{0}\approx-0.4855J/site$ for the $L=4$ cluster. The energy of the $L=2$ cluster is slightly higher than that obtained by the mVMC package\cite{mVMC}, which is $-0.5162J/site$\cite{Nikita}. However, we note that the mVMC result is obtained with additional quantum number projection procedure which we have not attempted\cite{mVMC,Nikita}. Such a quantum number projection procedure generates symmetric superposition of symmetry broken states. While such a symmetrization procedure can indeed lower the variational ground state energy on small clusters, it becomes ineffective in the thermodynamic limit, when it only results in a Schrodinger cat's state of the symmetry broken phase. 

On the $L=4$ cluster, our generalized RVB state outperforms the mVMC result, which is $E_{0}\approx-0.4831J/site$(see Fig.2). This is at first glance strange since the paffian wave function optimized in the mVMC package\cite{mVMC} contains our generalized RVB wave function as a subset. However, in the implementation adopted in Ref.[\onlinecite{Nikita}], the authors have assumed translational symmetry on the paffian matrix element $f_{i,j}$ to reduce the numerical complexity of the optimization procedure. Under such an assumption, the number of variational parameter is reduced from order $N^{2}$ to order $N$. On the other hand, in our unrestricted variational optimization of the generalized RVB ansatz we have made no assumption on the variational parameters. For example, on the $L=4$ pyrochlore cluster, the total number of variational parameter to be optimized is $N_{v}=4N^{2}=262144$. As we will see below, an unrestricted variational optimization is crucial to reveal the intricate nature of symmetry breaking pattern in the ground state of the spin-$\frac{1}{2}$ PAFH, which features an involved VBC pattern with an enlarged unit cell containing as many as $32$ sites. If we insist on imposing translational symmetry on the ansatz, such a fine structure in the ground state will be obscured.

\begin{figure}
\includegraphics[width=8cm]{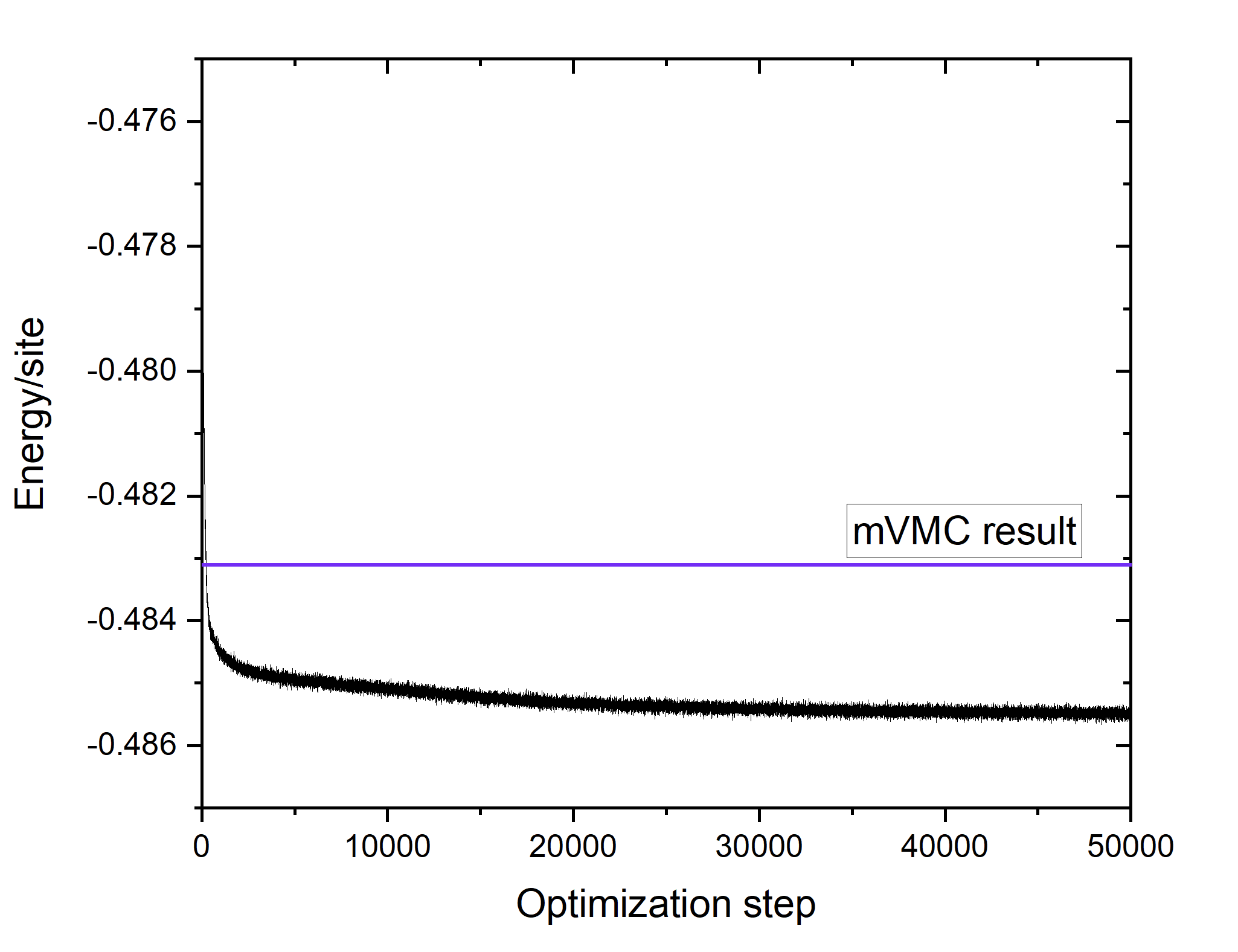}
\caption{Convergence of the variational energy of the spin-$\frac{1}{2}$ PAFH on a $L=4$ pyrochlore cluster with the optimization steps. Shown here is the result obtained from the unrestricted optimization of a generalized RVB ansatz which contains as many as $N_{v}=262144$ variational parameters. The purple horizontal line marks the energy obtained from the mVMC package with additional symmetry projection. Here we only show the evolution of the variational energy in the last 50000 optimization steps for brevity.}
\end{figure}

\subsection{Symmetry breaking pattern in the ground state of the spin-$\frac{1}{2}$ PAFH}

We find that the nature of the ground state of the spin-$\frac{1}{2}$ PAFH on the $L=2$ cluster is qualitatively different from that on the $L=4$ cluster. To tell such a difference we need an efficient way to diagnose the symmetry breaking pattern in the ground state. 

\subsubsection{$\langle \hat{\mathbf{S}}_{u}^{2}\rangle$ as a diagnosis of the symmetry breaking pattern in the ground state of the spin-$\frac{1}{2}$ PAFH}

The most elementary way to diagnose the symmetry breaking pattern in the ground state of the spin-$\frac{1}{2}$ PAFH is to compute the expectation value of the total spin squared on each tetrahedron of the pyrochlore lattice, namely $\langle \hat{\mathbf{S}}_{u}^{2}\rangle$. In the semiclassical limit of $S\rightarrow\infty$, the ground state of the PAFH is dictated by the ice rule of simultaneous vanishing of $\mathbf{S}_{u}$ on all tetrahedrons of the pyrochlore lattice. For the spin-$\frac{1}{2}$ PAFH, the ice rule can only be fulfilled locally but not globally. More specifically, the ice rule can be satisfied simultaneously on at most one half of all tetrahedrons, namely either on all the up-tetrahedrons or on all the down-tetrahedrons. Nevertheless, it is found by a recent pseudofermion functional renormalization group(PFFRG) study\cite{Iqbal2} that the bowtie structure in the static spin structure factor that is related to the ice rule remains approximately intact in the ground state of the spin-$\frac{1}{2}$ PAFH. It is thus of great interest to look into the expectation value of $\hat{\mathbf{S}}^{2}_{u}$ in the optimized ground state. 

For the spin-$\frac{1}{2}$ PAFH we have
\begin{equation}
\hat{\mathbf{S}}^{2}_{u}=\sum_{\substack{i\neq j\\i,j\in u}}\hat{\mathbf{s}}_{i}\cdot\hat{\mathbf{s}}_{j}+3
\end{equation}
We find that the expectation value of $\hat{\mathbf{S}}^{2}_{u}$ exhibits totally different behavior on pyrochlore cluster of linear size $L=2$ and $L=4$. For the $L=2$ cluster, we find that $\langle \hat{\mathbf{S}}^{2}_{u} \rangle$ is uniform on all tetrahedrons and 
\begin{equation}
\langle \hat{\mathbf{S}}^{2}_{u} \rangle=4E_{0}/J+3\approx 0.9528
\end{equation}
On the other hand, we find that $\langle \hat{\mathbf{S}}^{2}_{u} \rangle$ takes three different values on the $L=4$ pyrochlore cluster. More specifically, we have
\begin{equation}
\langle \hat{\mathbf{S}}^{2}_{u} \rangle \approx \left\{ 
\begin{aligned}
1.4552&, \ \ \ up \ tetrahedron\\ 
0.5804&, \ \ \  \frac{1}{4} \ of \ the \ down \ tetrahedron\\
0.6872&, \ \ \ \frac{3}{4} \ of \ the \ down \ tetrahedron\\
\end{aligned}\right.
\end{equation}
Obviously, the translational symmetry and the inversion symmetry relating the up and down tetrahedron are both broken on the $L=4$ cluster. As we will see below, such a symmetry breaking pattern is robust on larger pyrochlore clusters. 

\subsubsection{The local spin correlation pattern in the ground state of the $L=2$ and the $L=4$ pyrochlore cluster}

A finer diagnosis of the symmetry breaking pattern in the ground state of the spin-$\frac{1}{2}$ PAFH is provided by the expectation value of the spin correlation between nearest-neighboring site(the NN spin correlation). To elucidate the nature of the ground state of the spin-$\frac{1}{2}$ PAFH, we have computed the NN spin correlation on both the $L=2$ and the $L=4$ pyrochlore cluster. The results are illustrated in Fig.3 and Fig.4 respectively. 

For the $L=2$ cluster, we find that the NN spin correlation pattern remains translational symmetric. For this reason, we have only illustrated the result in a single pyrochlore unit cell. The ground state of the system is found to feature almost isolated spin chains along which the NN spin correlation is much stronger than that between different chains. As a result of the periodic boundary condition, each of these spin chains form a closed loop of length $4$. There are in total $8$ such spin chains in the $L=2$ pyrochlore cluster. If we neglect the spin correlation between different spin chains, then the ground state energy of system should be approximately $-0.5J/site$, which is the exact ground state energy of a spin-$\frac{1}{2}$ antiferromagnetic Heisenberg spin chain of length $4$ under periodic boundary condition. Such an estimation is very close to the optimized ground state energy we get on the $L=2$ pyrochlore cluster, which is $-0.5118J/site$. However, on larger pyrochlore clusters, such a quasi-one dimensional state may lose its stability. In fact, if we still use the above isolated spin chain approximation, then the ground state energy per site in the thermodynamic limit would be approximately $-\ln2 J+\frac{J}{4}\approx -0.4431J$, which is the exact ground state energy of an infinitely long spin-$\frac{1}{2}$ antiferromagnetic Heisenberg spin chain. As we will show below, this energy can be easily surpassed by other symmetry breaking phases on larger pyrochlore clusters.    

\begin{figure}
\includegraphics[width=8cm]{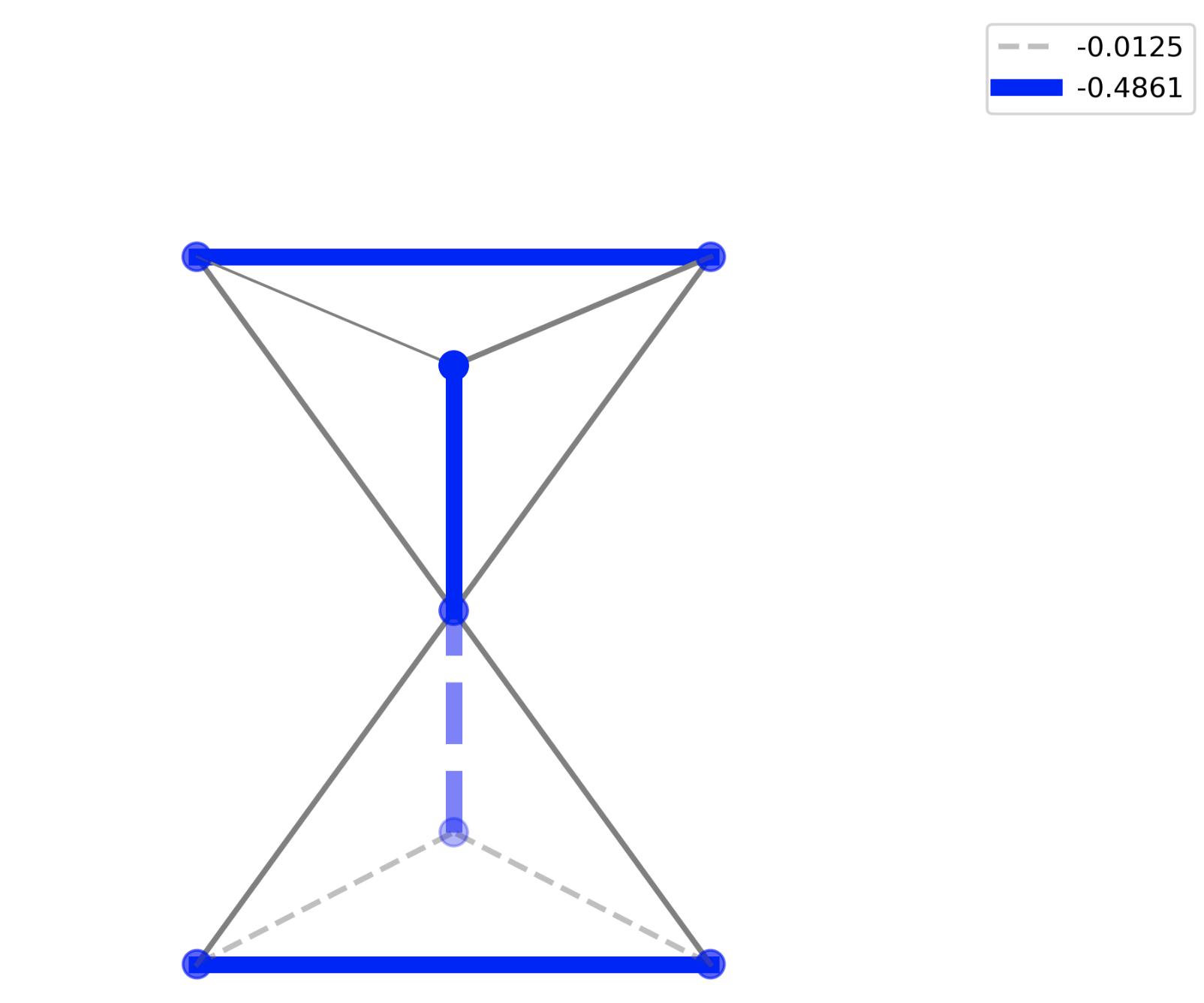}
\caption{The local spin correlation pattern in the ground state of the spin-$\frac{1}{2}$ PAFH on the $L=2$ pyrochlore cluster. Since the spin correlation pattern is translational symmetric, we have only illustrated the result in a single pyrochlore unit cell. Here we use the thickness of the bond to denote the strength of the NN spin correlation on it. The number beside the legend marks the expectation value of the spin correlation on the given bond. We note that the blue bonds actually form closed loops of length $4$ under periodic boundary condition.}
\end{figure}
  
The local spin correlation pattern for the $L=4$ pyrochlore cluster is much more complicated. The ground state of the system now features an intricate VBC pattern with $2\vec{a}_{1}\times2\vec{a}_{2}\times2\vec{a}_{3}$ periodicity(see Fig.4). The enlarged unit cell contains $32$ sites. There are at least four levels of hierarchical structure in such a VBC pattern. At the first level of the hierarchy, one find that all the down tetrahedrons are strongly spin dimerized. The spin dimerization is so strong that the spin correlation on the remaining $4$ weak bonds of the down tetrahedron is almost suppressed to zero. On the other hand, the NN spin correlation in the up tetrahedron is much more uniform. The inversion symmetry relating the up and the down tetrahedron is thus broken. As a result of such spin dimerization, the valence bonds in the system fall roughly into three categories, namely the strong valence bond in the down tetrahedron, the companying weak valence bond in the down tetrahedron, and the moderate valence bond in the up tetrahedron. Such strong spin dimerization is driven by the need to minimize the total spin squared in the down tetrahedrons. Indeed, the down tetrahedron has a $\langle\hat{\mathbf{S}}^{2}_{u}\rangle$ that is about three times smaller than that of the up tetrahedron. We note that the down and the up tetrahedron can not be simultaneously strongly dimerized.

A closer inspection of Fig.4 reveals the second level of hierarchical structure in the VBC pattern. More specifically, among all the $N_{c}$ down tetrahedrons, $\frac{1}{4}$ of the down tetrahedron have stronger dimerization than the remaining $\frac{3}{4}$ down tetrahedron. The unit cell is doubled in all the three basis vector directions of the pyrochlore lattice. For sake of clearance, we have illustrated the first and the second layer of the enlarged unit cell in the $\vec{a}_{3}$ direction separately in Fig.4. These $\frac{1}{4}$ more strongly dimerized down tetrahedrons also have a slightly smaller $\langle\hat{\mathbf{S}}^{2}_{u}\rangle$. Such a finer structure is driven by the $6$-spin resonance process around the hexagonal rings as illustrated in Fig.4 by blue shaded areas. These hexagonal rings are made up of three strong valence bonds in the less strongly dimerized down tetrahedrons(the blue bond) and three moderate bonds in the up tetrahedrons(the cyan bond). We note that on each kagome plane of the pyrochlore cluster of linear size $L$, there are at most $\frac{L^{2}}{4}$ such resonating rings(see Fig.4c). This can be seen by noting the fact that the centers of the hexagon in a kagome plane form a triangular lattice and that neighboring hexagon in the kagome plane can not participate in the $6$-spin resonance process simultaneously. The pyrochlore lattice hosts four set of kagome planes directed respectively in the $[1,0,0]$, $[0,1,0]$, $[0,0,1]$ and the $[1,1,1]$ direction. Thus there are at most $L^{3}$ such resonating rings on the pyrochlore cluster of linear size $L$. The VBC pattern we obtained from the unrestricted optimization saturates this number. We thus call such a VBC pattern the maximally resonating VBC pattern. Obviously, the $2\vec{a}_{1}\times2\vec{a}_{2}\times2\vec{a}_{3}$ periodicity of the VBC pattern is dictated by such a maximally resonating condition. At the same time, the stronger dimerization in the $\frac{1}{4}$ down tetrahedrons that do not participate in the $6$-spin resonance process can be attributed to the more localized nature of these valence bonds(the black bond).   

\begin{figure}
\includegraphics[width=8cm]{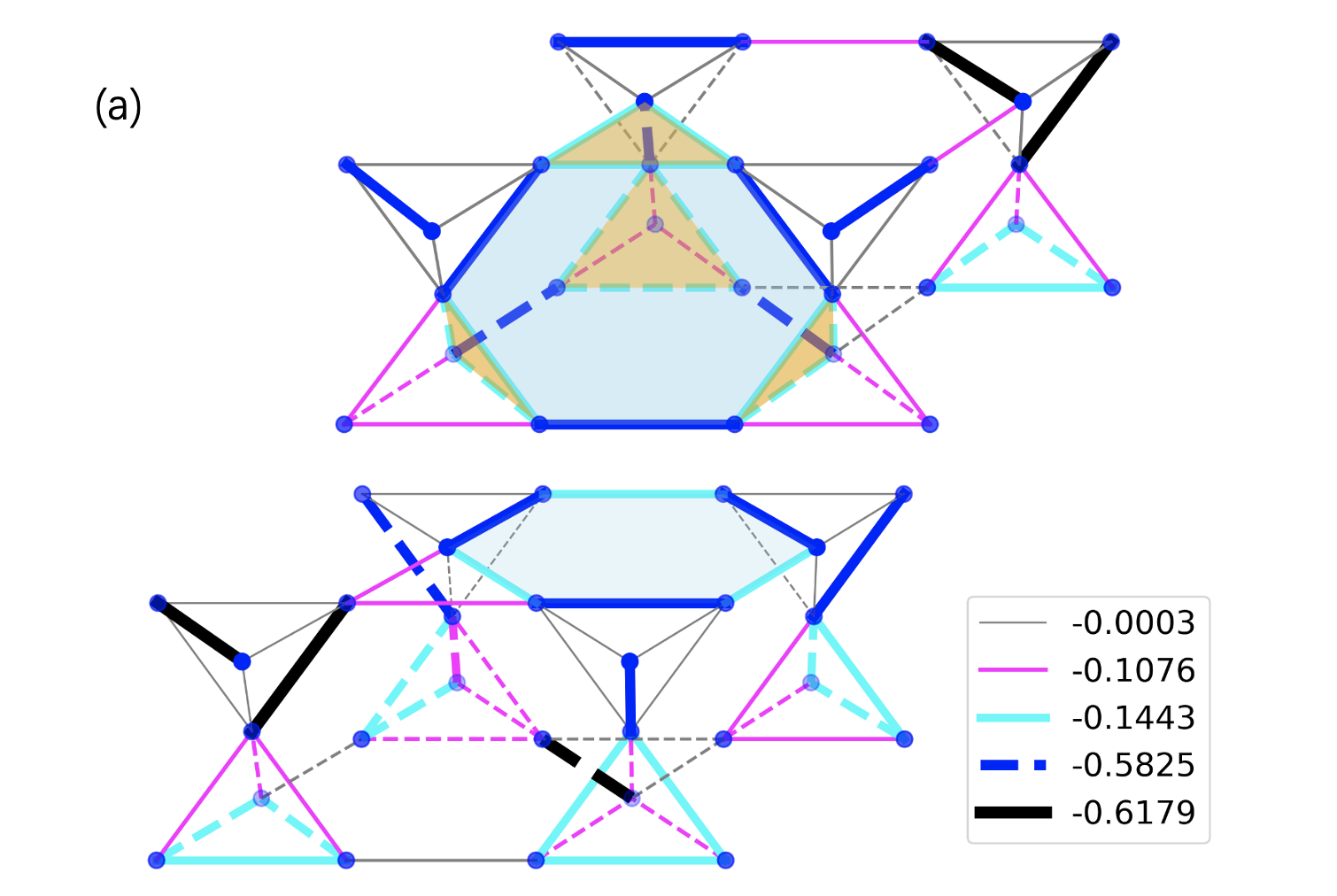}
\includegraphics[width=8cm]{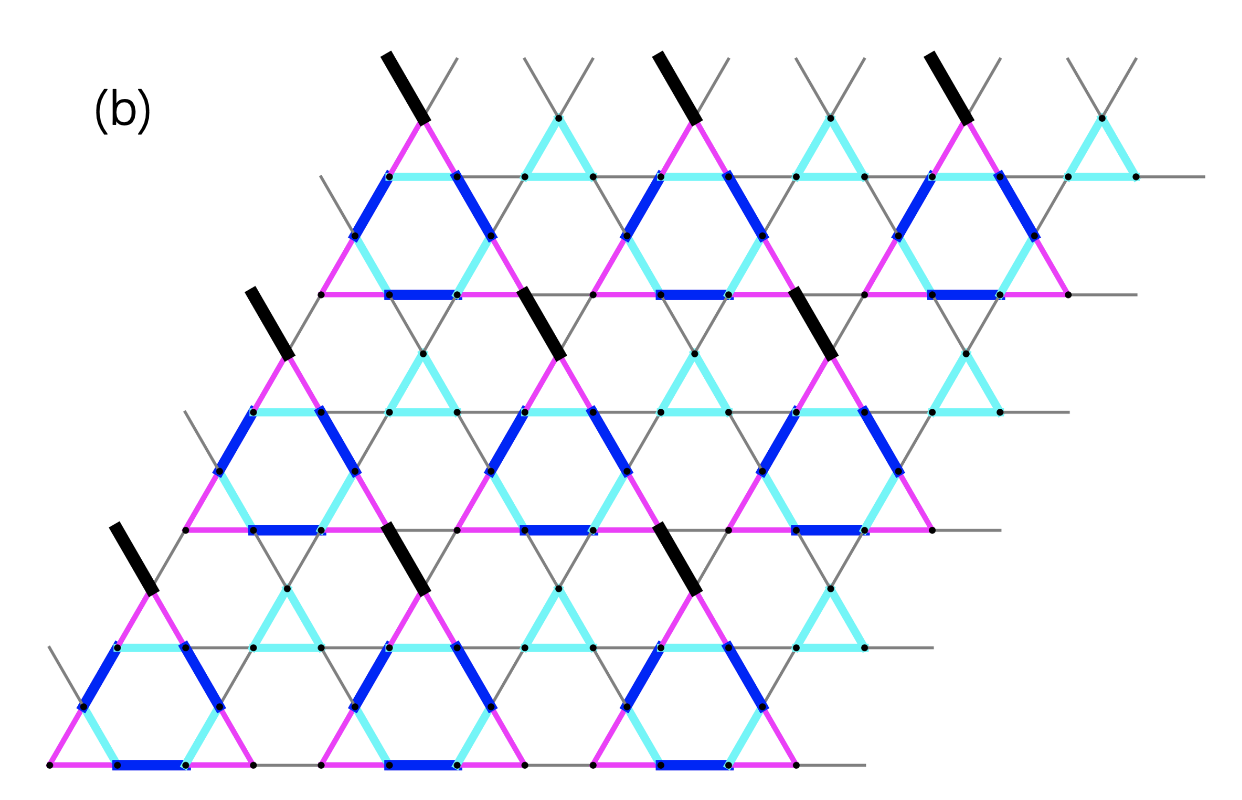}
\caption{The local spin correlation pattern in the maximally resonating VBC state. The unit cell is doubled in all the three basis vector directions of the pyrochlore lattice and thus contains in total $32$ lattice sites. For sake of clearance, we have illustrated the two layers of the enlarged unit cell in the $\vec{a}_{3}$ direction separately. Here we use both color and thickness of the bond to distinguish valence bonds of different strength. The number beside the legend denotes the expectation value of the NN spin correlation on the given bond. The blue shaded hexagonal areas mark the location of the $6$-spin resonance process. (c)Illustration of the arrangement of the resonating rings on a kagome plane of pyrochlore lattice in the maximally resonating VBC state. We note that there are four set of kagome planes in the pyrochlore lattice, directed respectively in the $[1,0,0]$, $[0,1,0]$, $[0,0,1]$ and the $[1,1,1]$ direction.}
\end{figure}

Beside these two base hierarchical levels, there are still higher level of hierarchical structures in the symmetry breaking pattern of the maximally resonating VBC state. To see this, we note that the spin dimerization pattern in the $\frac{1}{4}$ most strongly dimerized down tetrahedrons remain undecided even if we take into account the $6$-spin resonance process around the hexagonal rings. More specifically, there are $3$ different ways to arrange the two strongest valence bonds(the black bond in Fig.4) within each such down tetrahedron. The degeneracy related to this ambiguity is thus $3^{\frac{N_{c}}{4}}$ on a pyrochlore cluster with $N_{c}$ unit cell. To lift this degeneracy, we should take into account higher order spin resonance processes. Such higher order spin resonance processes contribute very little to the variational ground state energy. However, our finite-depth BFGS algorithm can still successfully capture such a subtle detail.      
 
These three levels of hierarchical structure however still do not exhaust all structure hierarchy of the maximally resonating VBC state, since what we have done only concerns the arrangement of the strongest valence bonds within the down tetrahedrons. When we consider the arrangement of the moderate valance bonds within the up tetrahedrons, we will find more hierarchy levels. The structure at these higher level of hierarchy contributes even less to the ground state energy. Nevertheless, our unrestricted variational optimization can reveal clearly the structure in the arrangement of the moderate valence bonds. More specifically, the moderate valence bonds within the up tetrahedron fall again into two subgroups. Within each up tetrahedron, there are three cyan bonds with slightly stronger NN spin correlation and three pink bonds with slightly weaker NN spin correlation. The three pink bonds always meet at a vertex that is shared by a down tetrahedron that does not participate in the $6$-spin resonance process. The cyan bonds with the larger NN spin correlation are always involved in the $6$-spin resonance process. 

Finally, we note that the result shown in Fig.4 is actually obtained on a $L=6$ pyrochlore cluster. The spin correlation pattern on the $L=4$ cluster has a slightly lower symmetry as the cyan bonds split further into two subgroups with slightly different spin correlation. We think that such an additional symmetry breaking on the $L=4$ cluster is a finite size effect.   
 
 \subsubsection{An isolated spin chain state or a maximally resonating VBC state?}
 We note that both the isolated spin chain state found on the $L=2$ cluster and the maximally resonating VBC state found on the $L=4$ cluster are compatible with the periodic boundary condition imposed on both clusters. We thus must understand why the isolated spin chain state loses its stability on the $L=4$ cluster, or vice versa, why the maximally resonating state is not as competitive as the isolated spin chain state on the $L=2$ cluster. As we have mentioned above, an ideal isolated spin chain state on the $L=2$ cluster has the energy of $-0.5J/site$. This is very close to the optimized ground state energy we found on such a cluster, which is $-0.5118J/site$. In the thermodynamic limit, the ideal isolated spin chain state has an energy of $-0.4431J/site$. This is much higher than the optimized ground state energy we find on the $L=4$ cluster, which is $-0.4855J/site$. To check if the isolated spin chain state is as competitive on the $L=4$ cluster, we have computed the ground state energy of a length-$8$ spin-$\frac{1}{2}$ antiferromagnetic Heisenberg spin chain with periodic boundary condition. We find the result to be $-0.4564J/site$. This is again significantly higher than the optimized ground state energy we find on the $L=4$ cluster. We thus think the dominance of the isolated spin chain state on the $L=2$ cluster is purely a finite size effect. Indeed, as a quasi one dimensional state, the isolated spin chain state has a much stronger finite size effect in its spin correlation than the three dimensional maximally resonating VBC state. This explains why the maximally resonating VBC state fails to be the ground state of the $L=2$ cluster, although it is also compatible with the periodic boundary condition on such a cluster.

 \subsubsection{NN-RVB ansatz for the maximally resonating VBC state and its extension on larger pyrochlore clusters}
The same unrestricted optimization using the generalized RVB ansatz can certainly be carried out on larger pyrochlore clusters. However, since the number of variational parameters grows with the linear size of the system as $64\times L^{6}$, this is numerically expensive. For example, on a $L=6$ pyrochlore cluster the number of variational parameters is $N_{v}=2985984$. This number increases further to $N_{v}=16777216$ for a $L=8$ pyrochlore cluster. Although the finite-depth BFGS algorithm still works well in such situations, an unrestricted optimization of the generalized RVB ansatz from totally random initial guess is inefficient. We'd better start from a more educated guess. For this purpose, we will first attempt a more restrictive ansatz which has better scalability and demands less computational resources.   
 
 Here we will approximate the ground state of the spin-$\frac{1}{2}$ PAFH with NN-RVB ansatz of the following form
 \begin{eqnarray}
H_{\mathrm{MF}}&=&-\sum_{\langle i,j \rangle,\alpha}[\chi_{i,j} f^{\dagger}_{i,\alpha}f_{j,\alpha}+\Delta_{i,j} f^{\dagger}_{i,\alpha}f^{\dagger}_{j,\bar{\alpha}}]+h.c.\nonumber\\
&+&\sum_{i,\alpha}\mu_{i}f^{\dagger}_{i,\alpha}f_{i,\alpha}
\end{eqnarray}
Since the maximally resonating VBC pattern features a $2\vec{a}_{1}\times2\vec{a}_{2}\times2\vec{a}_{3}$ periodicity, we will assume the same $2\vec{a}_{1}\times2\vec{a}_{2}\times2\vec{a}_{3}$ periodicity on the variational parameter $\chi_{i,j}$, $\Delta_{i,j}$ and $\mu_{i}$. We note that the translational symmetry in a RVB state can in general be realized projectively on the RVB mean field ansatz as a result of the gauge redundancy in the slave particle representation of the spin operator. However, we find that the optimized NN-RVB ansatz with $2\vec{a}_{1}\times2\vec{a}_{2}\times2\vec{a}_{3}$ periodicity can reproduce every qualitative feature of the maximally resonating VBC state we obtained from unrestricted optimization of the generalized RVB ansatz. It is thus feasible to impose such a periodicity on the RVB mean field ansatz. At the same time, we find that the time reversal symmetry is unbroken in the ground state. It is thus permissible to restrict the optimization to real ansatz. With these considerations in mind, we are left with $N_{v}=224$ parameters to be optimized in the NN-RVB ansatz. More specifically, we have $96$ hopping parameter $\chi_{i,j}$, $96$ pairing parameter $\Delta_{i,j}$ and $32$ local chemical potential $\mu_{i}$ in the NN-RVB ansatz.  The energy gradient with respect to these $N_{v}=224$ parameters can be computed with the first order perturbation theory method introduced in Sec.II.

Since the generalized RVB ansatz contains the NN-RVB ansatz as a subset, the variational energy we get from optimizing the NN-RVB ansatz should be higher than that obtained from optimizing the generalized RVB ansatz. This is indeed the case. On the $L=4$ pyrochlore cluster, we find that the optimized variational energy for the NN-RVB ansatz is $-0.4825J/site$(see Fig.5). We can then use the optimized solution as an initial guess to optimize the NN-RVB ansatz on the $L=6$ cluster. The solution can again be used as an initial guess to optimize the NN-RVB ansatz on the $L=8$ cluster. We find that the variational ground state energy of the $L=6$ cluster converges to almost the same value as the $L=4$ cluster, namely $-0.4825J/site$(see Fig.5a). The finite size effect of the variational energy calculated form the NN-RVB ansatz is thus very small on pyrochlore cluster with $L\ge 4$. On the $L=8$ cluster, we find that the variational ground state energy converges to slightly lower value of $-0.48266J/site$(Fig.5b). In addition, we find that the maximally resonating VBC pattern remains intact when we increase the linear system size.

\begin{figure}
\includegraphics[width=8cm]{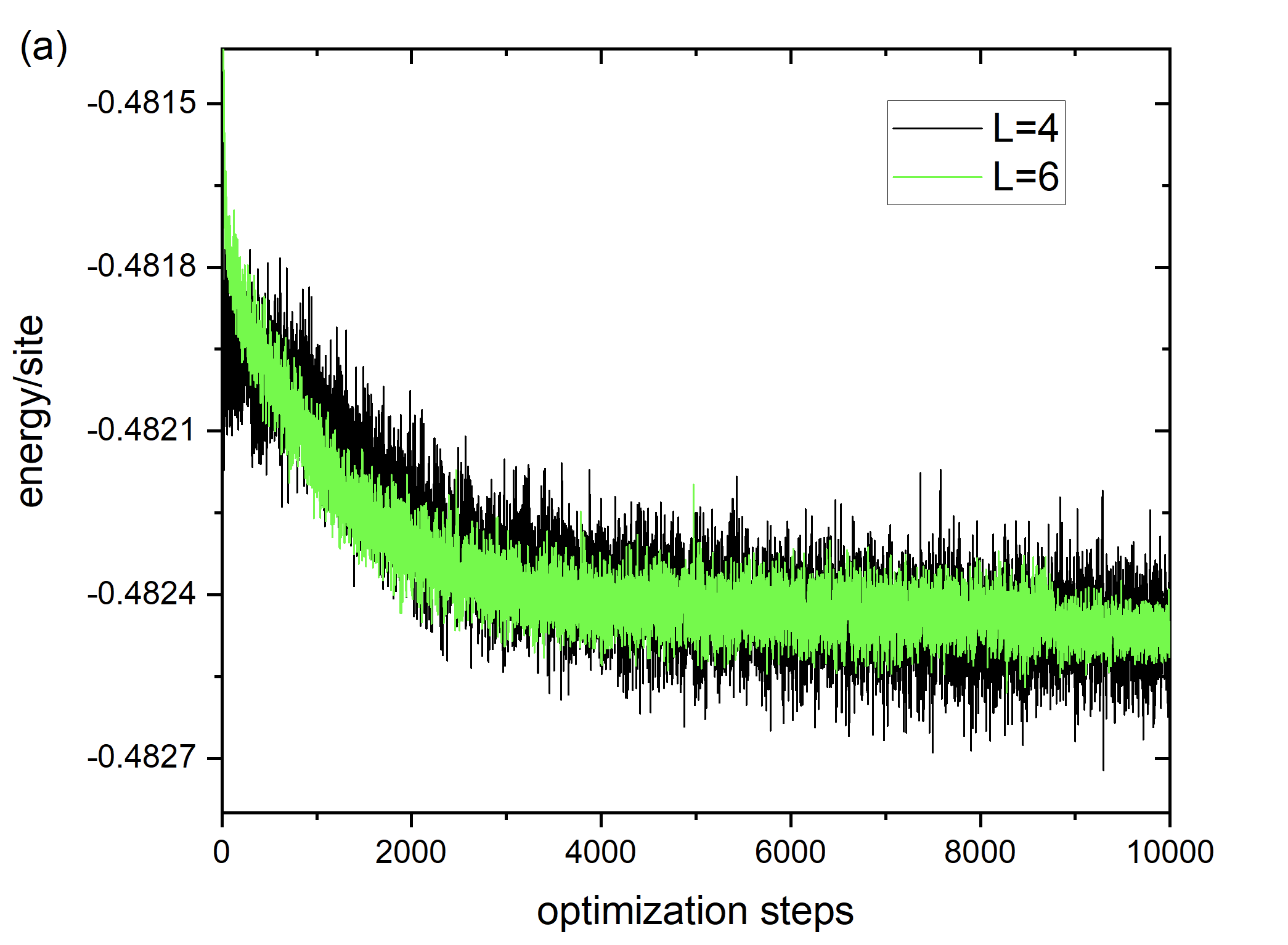}
\includegraphics[width=8cm]{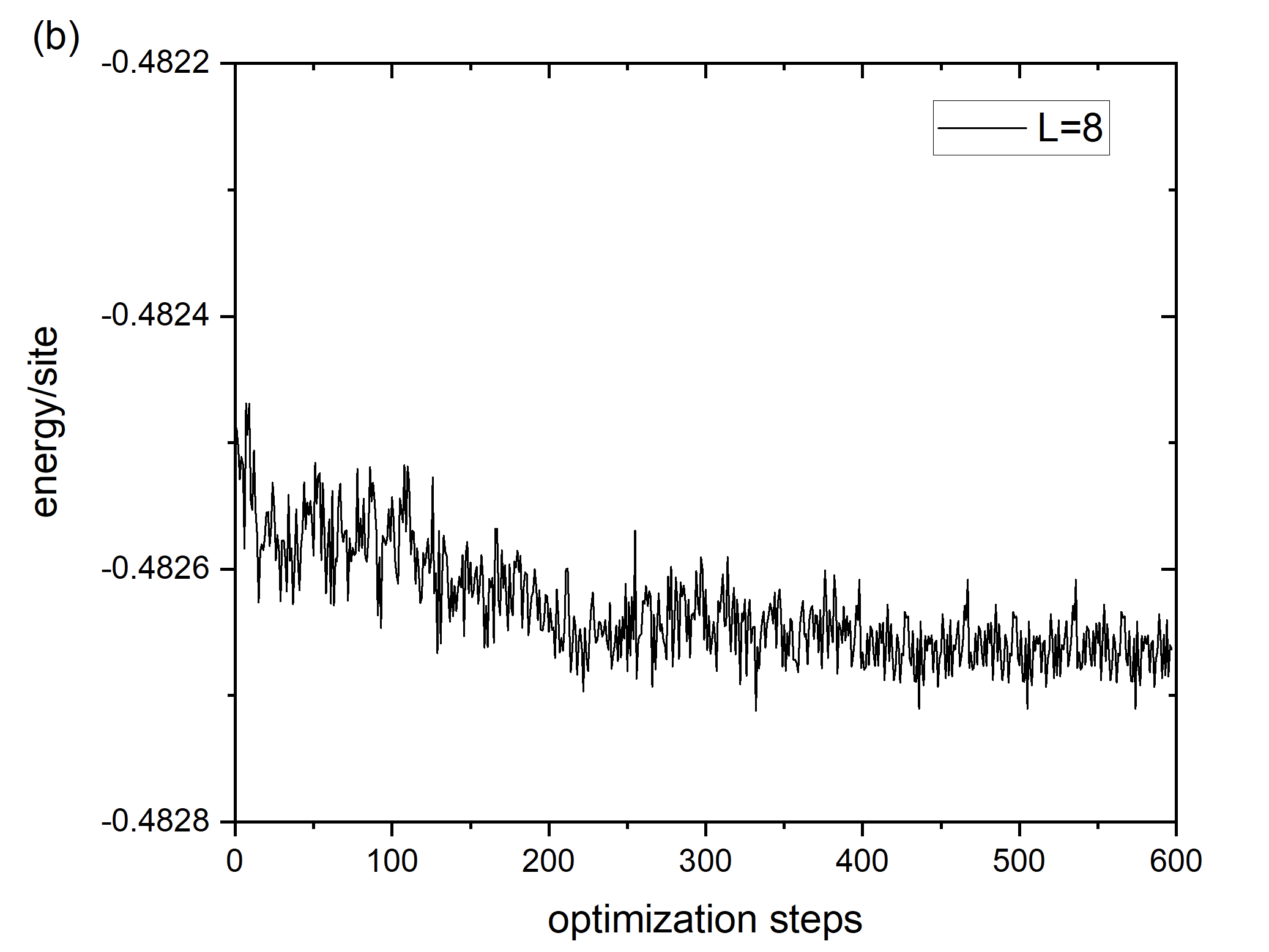}
\caption{The convergence of the variational energy of the spin-$\frac{1}{2}$ PAFH with optimization steps on the $L=4$, $L=6$ and the $L=8$ pyrochlore cluster. The results are obtained by optimizing the NN-RVB ansatz. Here we use the optimized NN-RVB ansatz on the $L=4$ cluster as the initial guess for the $L=6$ cluster. The optimized ansatz for the $L=6$ cluster is again used as the initial guess for the $L=8$ cluster. For both the $L=4$ and the $L=6$ cluster, we have shown here the evolution of the variational energy during the last 10000 optimization steps. For the $L=8$ cluster, we have only conducted 600 optimization steps as a result of the rapid convergence of the variational energy itself and the heavy computational cost on such a large cluster.}
\end{figure}

Beside being a huge simplification of the generalized RVB ansatz, the optimized NN-RVB ansatz can also be used as an educated initial guess for unrestricted optimization of the generalized RVB ansatz. We have attempted this procedure on both the $L=6$ cluster and the $L=8$ cluster. We note that the number of the variational parameters to be optimized is as large as $N_{v}=16777216$ on the $L=8$ cluster. An unrestricted optimization starting from random initial guess is too expensive for us to afford. However, as can be seen from Fig.6, the finite-depth BFGS algorithm still works well if we adopt the optimized NN-RVB as the initial guess. The converged variational ground state energy are $-0.4848J/site$ for the $L=6$ cluster and $-0.4847J/site$ for the $L=8$ cluster.  

\begin{figure}
\includegraphics[width=8cm]{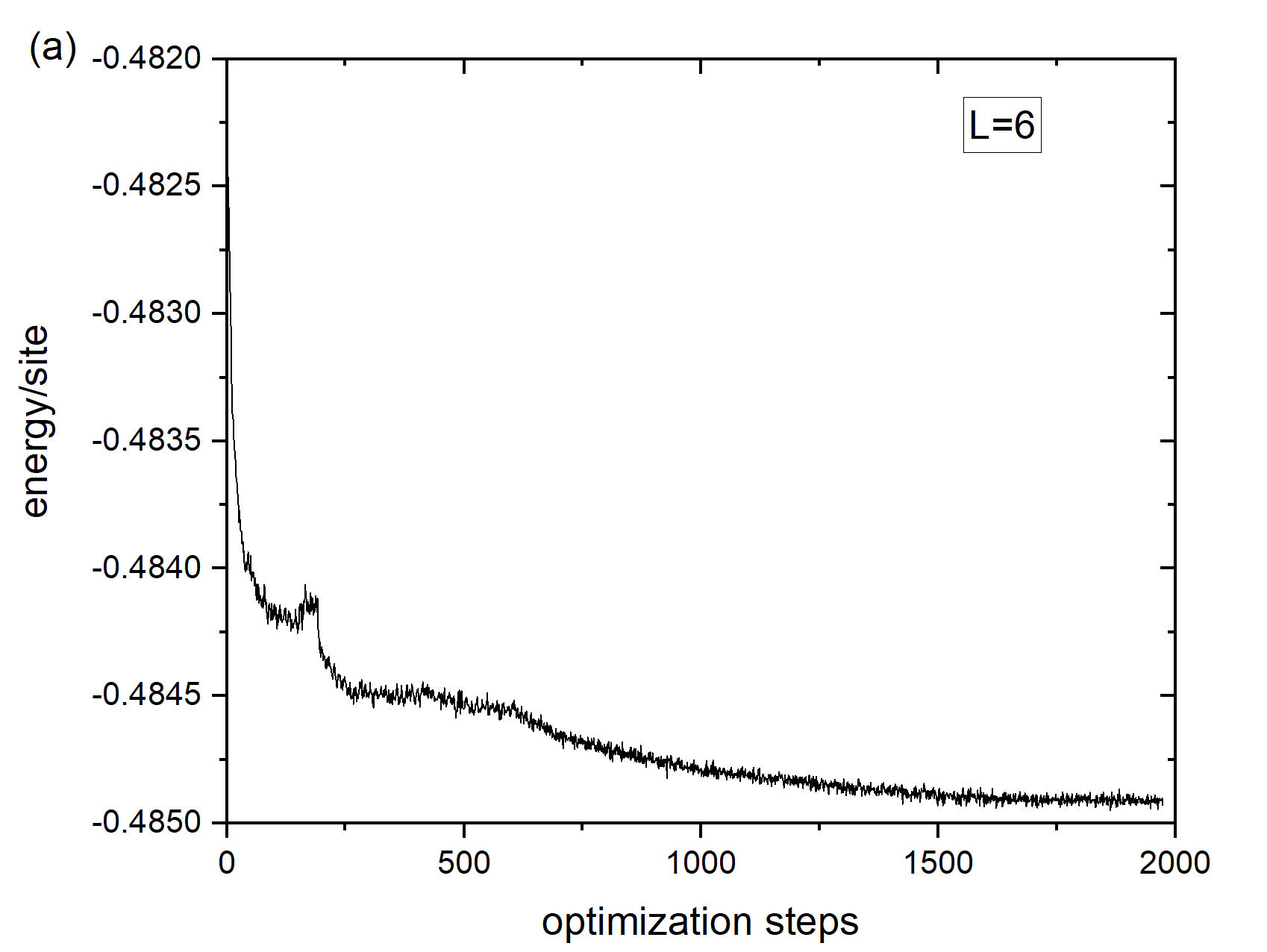}
\includegraphics[width=8cm]{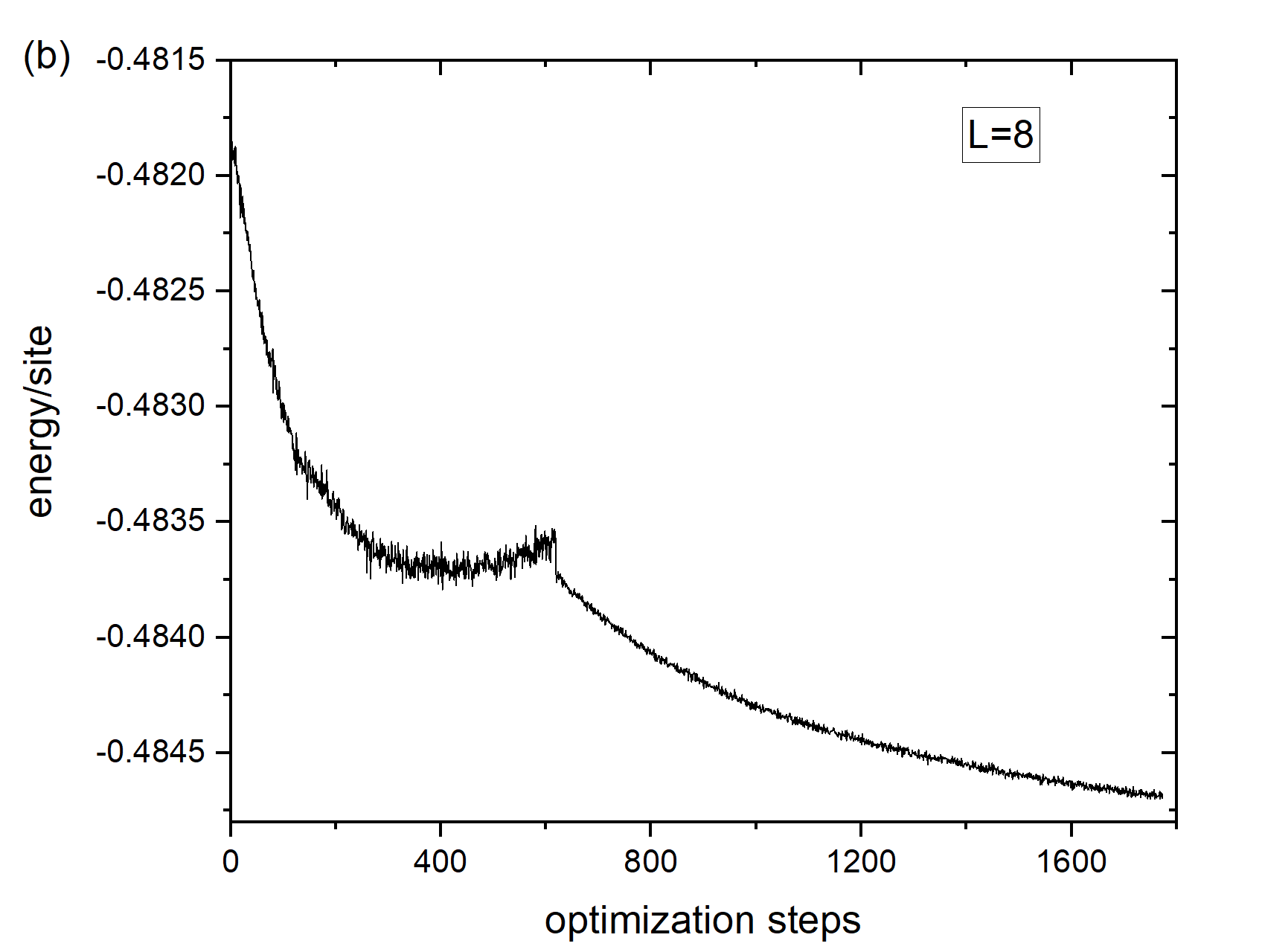}
\caption{The convergence of the variational energy of the spin-$\frac{1}{2}$ PAFH with optimization steps on the $L=6$ and the $L=8$ pyrochlore cluster. The results are obtained by optimizing the generalized RVB ansatz. Here we use the optimized NN-RVB ansatz on the same cluster as the initial guess for the unrestricted optimization of the generalized RVB ansatz. Note that there are $N_{v}=2985984$ variational parameters on the $L=6$ cluster and $N_{v}=16777216$ variational parameters on the $L=8$ cluster. The finite-depth BFGS algorithm behaves very well even in such extreme situations. To save computational resource we have stopped the optimization when we find that the last 100 optimization steps fail to decrease the variational energy significantly further. The sudden change of slope in the energy data is caused by resetting of optimization parameters for a finer search.}
\end{figure}   

 In principle, the same procedure can be extended to larger clusters. However, in the calculation of the energy gradient of the NN-RVB state we need the information of $\nabla \mathbf{U}$, the storage demand of which scales as $N_{v}\times N^{2}$ with the system size. For a $L=10$ cluster, which contains $N=4000$ lattice site, the storage demand is $27$Gb. Our mid-sized optimization project can not afford such a large storage demand. Of course, we can distribute the information contained in $\nabla \mathbf{U}$ in many processors or compute it on-the-fly. However, these alternatives all require larger computational resources that are currently unavailable to us.     
Nevertheless, we find that the energy data obtained on systems up to $L=8$ is already sufficient for us to make meaningful extrapolation to the thermodynamic limit. In Fig.7, we compare the size scaling of the variational ground state energy obtained in this work and that obtained from previous works. The extrapolated ground state energy is $-0.4827J/site$ for the NN-RVB ansatz and $-0.4846J/site$ for the generalized RVB ansatz. These results are well below previous estimate($-0.477J/site$) based on the RVB theory and constitute new benchmarks for the spin-$\frac{1}{2}$ PAFH.

\begin{figure}
\includegraphics[width=8cm]{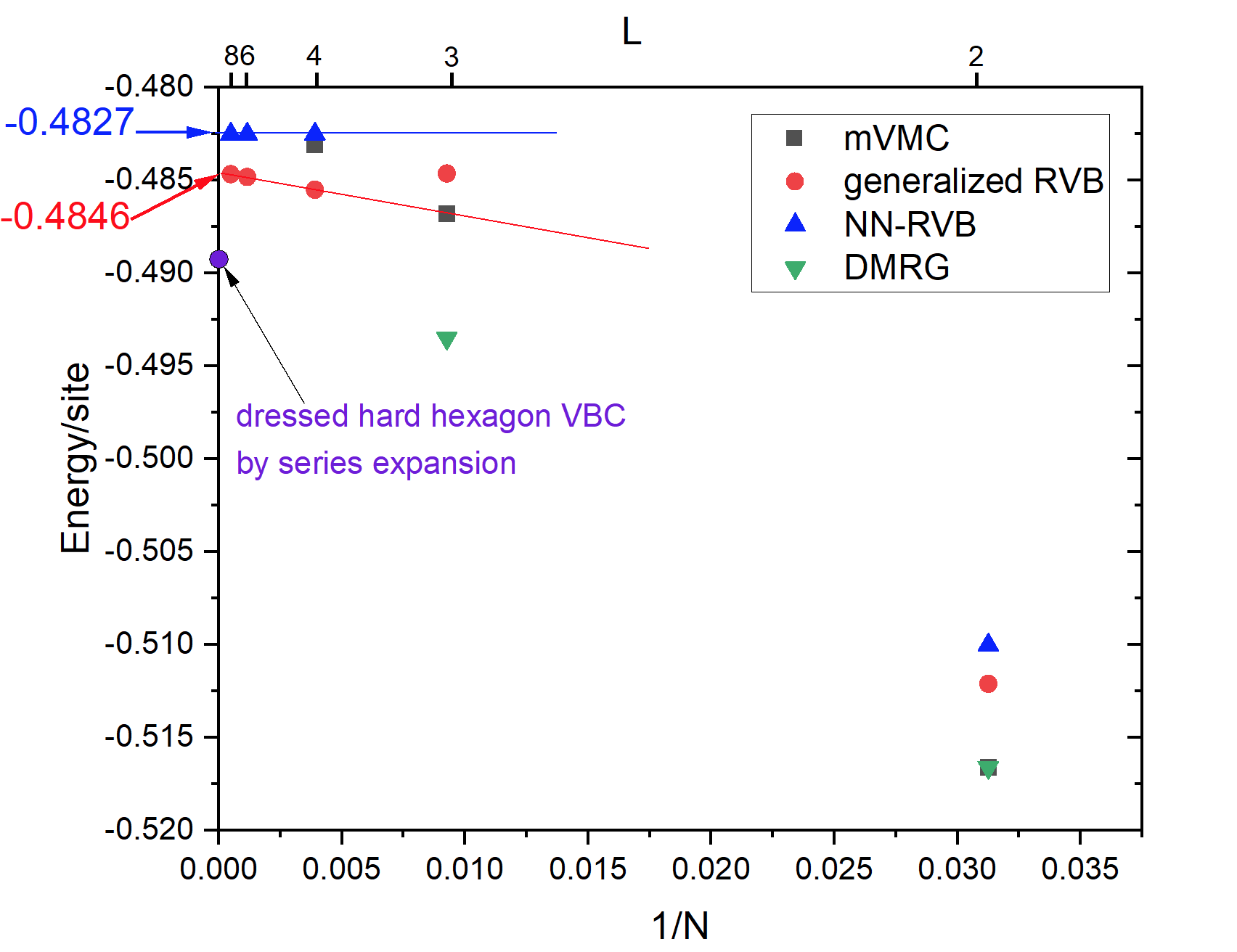}
\caption{Comparison of the variational ground state energy of the spin-$\frac{1}{2}$ PAFH obtained with different approaches. Here we only compare the results obtained on equilateral pyrochlore cluster with the $L\times L\times L\times 4$ geometry. $N=4L^{3}$ denotes the number of lattice site in the cluster. The red and blue line mark the linear fit of the variational energy obtained from optimizing the generalized RVB ansatz and the NN-RVB ansatz. In the linear fit we have only included the result obtained on clusters with linear size $L=4,6,8$ since the $L=3$ cluster is not compatible with the $2\vec{a}_{1}\times 2\vec{a}_{2}\times 2\vec{a}_{3}$ periodicity of the maximally resonating VBC pattern and that an isolated spin chain state is favored on the $L=2$ cluster. Here the black symbol denotes the result obtained from the mVMC package in Ref.[\onlinecite{Nikita}], green symbol denotes the result obtained from the DMRG simulation in Ref.[\onlinecite{Luitz}]. The purple dot on the vertical axis denotes the energy of the dressed hard hexagon VBC state calculated in the thermodynamic limit by series expansion on the dressing parameter[\onlinecite{Robin}].}
\end{figure}   

 \subsubsection{A comparison with the dressed hard hexagon VBC state}
Very recently, a VBC state obtained from dressing hard hexagon covering of the pyrochlore lattice is found to have a very good variational energy. To construct such a state, we first decompose the pyrochlore lattice into non-overlapping hexagonal rings so that each site is involved in one and only one such hexagonal ring. Clearly, there is $N/6$ such hexagonal rings on a pyrochlore cluster with $N$ sites. The dressed hard hexagon VBC state is exponentially numerous in the linear size of the system. For a given hard hexagon covering we decompose the model Hamiltonian as follows
\begin{equation}
H=H_{0}+V
\end{equation}
Here $H_{0}$ denotes the Heisenberg exchange coupling between NN spins belonging to the same hexagon, $V$ denotes the Heisenberg exchange coupling between NN spins that belongs to different hexagons. If we neglect $V$, then the ground state of the system is given by the direct product of the ground state of each hexagonal rings, whose energy is $-0.4671J/site$. The dressed hard hexagon VBC state is then constructed as follows
\begin{equation}
|\Psi_{\alpha}\rangle=e^{-\alpha V}|\Psi_{0}\rangle
\end{equation}
Here $|\Psi_{0}\rangle$ denotes the direct product of the ground state of the hexagonal rings.  

Through series expansion on the dressing parameter $\alpha$, the energy of the dressed hard hexagon VBC state is found to be $-0.48947J/site$ in the thermodynamic limit. This sets the lowest known upper bound on the ground state energy of the spin-$\frac{1}{2}$ PAFH. We note that the structure of this VBC state is drastically different from the maximally resonating VBC state found in this work. For example, $\langle\hat{\mathbf{S}}_{u}^{2}\rangle$ remains uniform on all tetrahedrons in $|\Psi_{\alpha}\rangle$. In fact, from the construction we see that $|\Psi_{\alpha}\rangle$ has exactly the same symmetry as the bare hard hexagon covering of the pyrochlore lattice. Although inversion, rotation and translation symmetry are all broken in such a VBC state, it is invariant under the combined operation of spatial inversion and translation or rotation. As a result, all tetrahedrons on the pyrochlore lattice are related to each other by symmetry operations. This is totally different from the maximally resonating VBC state found in this work, in which $\langle\hat{\mathbf{S}}_{u}^{2}\rangle$ on up and down tetrahedrons are strongly different. 
 
Naively, one would conclude that the dressed hard hexagon VBC state is more stable than the maximally resonating VBC state found in this work. However, we note that to decide the relative stability of two variational states, we must compare their energy computed at the same approximation level and under the same condition. After all, the RVB wave function is constructed from Gutzwiller projection of a single Slater determinant. With such a huge simplicity in the form of the wave function structure, we do not expect the RVB wave function to be quantitatively as competitive as other numerical methods in predicting the ground state energy of a system. In addition, as $|\Psi_{\alpha}\rangle$ has exactly the same symmetry as the bare hard hexagon covering, which is exponentially numerous in the linear size of the system, it is not at all clear if additional symmetry breaking would occur under further unrestricted optimization. 

With these considerations in mind, we have performed variational optimization on pyrochlore cluster that is large enough to accommodate both the dressed hard hexagon VBC state and the maximally resonating VBC state. It is easy to find that to accommodate both types of VBC state an equilateral pyrochore cluster must have a linear size of multiples of $6$, namely $L=6n$. Thus the minimal equilateral pyrochlore cluster that can meet such a requirement contains $N=4\times 6^{3}=864$ sites. For ease of implementation, in the following we will focus on a hard hexagon covering with $2\vec{a}_{1}\times3\vec{a}_{2}\times2(\vec{a}_{3}-\vec{a}_{2})$ periodicity as illustrated in Fig.9a. Other coverings, such as the one illustrated in Ref.[\onlinecite{Robin}], lead to essentially the same conclusion. 

To obtain the dressed hard hexagon VBC state, we start our optimization from an $U(1)$ RVB mean field ansatz of the following form
\begin{equation}
H_{MF}=-\sum^{6}_{r,\alpha,i=1}f^{\dagger}_{i,r,\alpha}f_{i+1,r,\alpha}
\end{equation}
in which $f_{i,r,\alpha}$ denotes the spinon operator on the $i$-th site of the $r$-th ring in the hard hexagon covering. Here we have assumed periodic boundary condition on each hexagonal ring such that $f_{i+6,r,\alpha}=f_{i,r,\alpha}$. RVB mean field ansatz of this form is known to be extremely accurate for the spin-$\frac{1}{2}$ Heisenberg chain. For the $6$-spin ring in the hard hexagon covering state, the variational energy generated from such an $U(1)$ ansatz is $-0.4667J/site$. This is very close to the exact ground state energy of the length-$6$ spin-$\frac{1}{2}$ Heisenberg ring.  

\begin{figure}
\includegraphics[width=8cm]{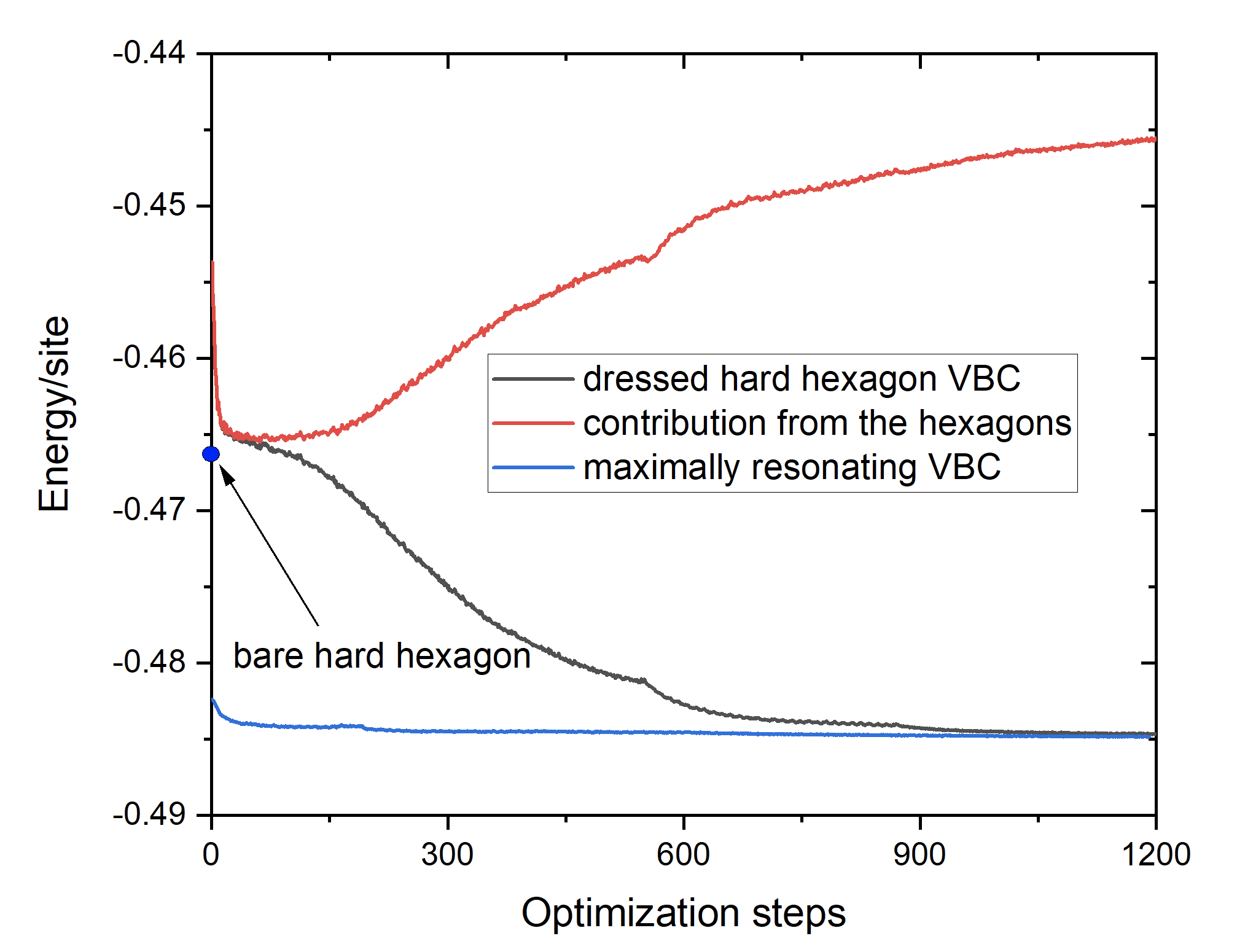}
\caption{Comparison of the convergence of the variational energy of the spin-$\frac{1}{2}$ PAFH for the maximally resonating VBC state(blue line) and the dressed hard hexagon VBC state(black line) on the $L=6$ cluster. These results are obtained by optimizing the generalized RVB ansatz. For the dressed hard hexagon VBC state, we have used the mean field ansatz Eq.42 with small random deviation as the initial guess for the unrestricted optimization. The red line illustrates the evolution of the contribution to the ground state energy from the hexagonal rings in the dressed hard hexagon VBC state. The blue dot on the vertical axis marks the energy of a bare hard hexagon state, which is $-0.4667J/site$ in the RVB approximation and $-0.4671J/site$ from exact calculation.The sudden change of slope in the energy data is caused by resetting of optimization parameters.}
\end{figure}   

The evolution of the variational energy on the $L=6$ cluster is illustrated in Fig.8. Here we have introduced a small and random deviation from the $U(1)$ ansatz given by Eq.42 in the initial guess to ensure that the optimization procedure can safely escape from the bare hard hexagon state, which is itself a saddle point of the variational energy. Intriguingly, we find that the energy of the dressed hard hexagon VBC state converges to a value that is almost indistinguishable from that of the maximally resonating VBC state. More specifically, we find the energy difference between the two states is smaller than $10^{-4}J/site$. At the same time, while the contribution to the total energy from the hexagonal rings decreases monotonically with optimization in the dressed hard hexagon state, it is never below $90\%$. This implies that the hard hexagon structure remains robust in the dressed state. To see this more clearly, we have calculated the local spin correlation in the fully optimized state. The result is illustrated in Fig.9b.  

\begin{figure}
\includegraphics[width=8cm]{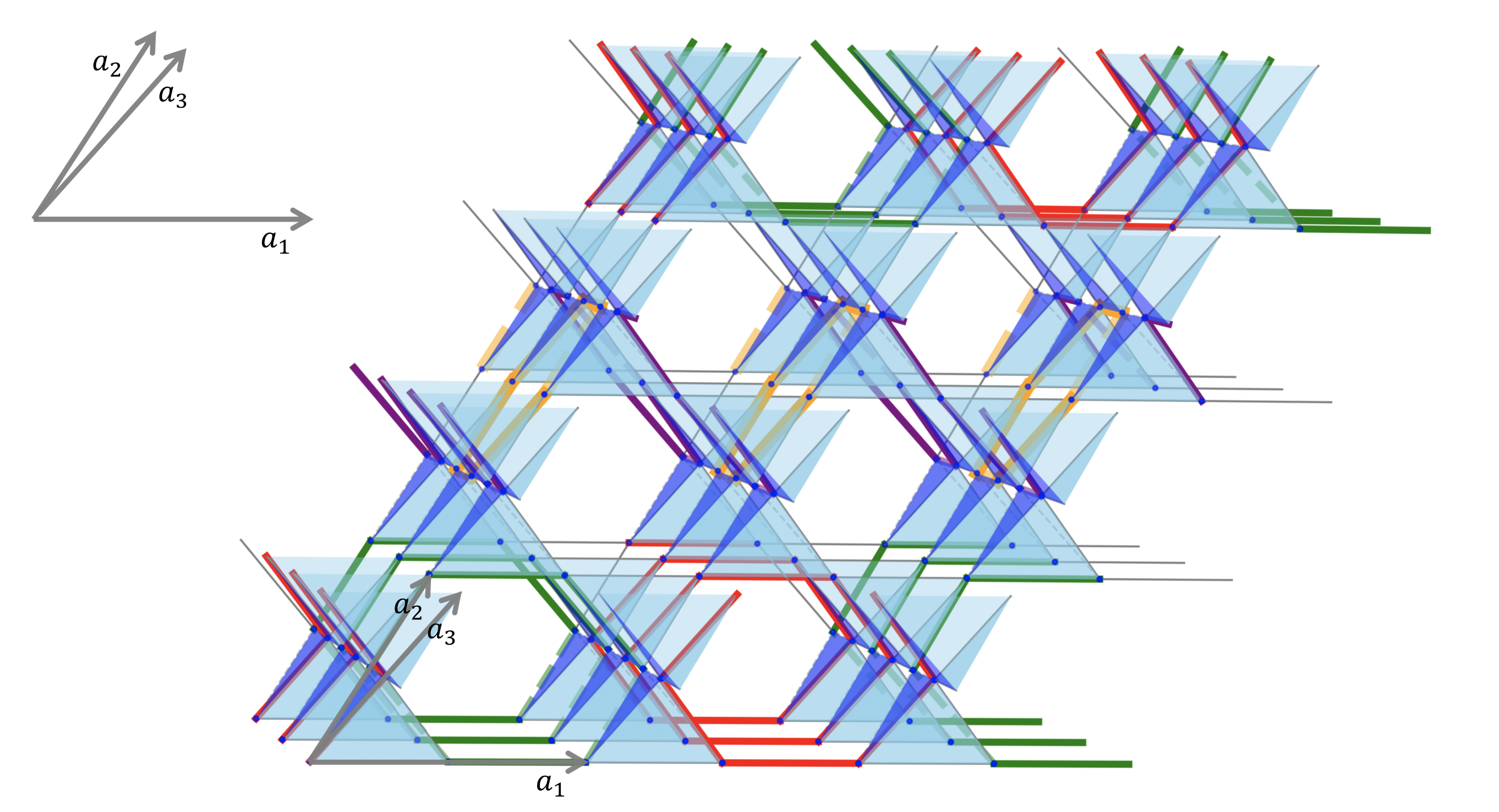}
\includegraphics[width=9cm]{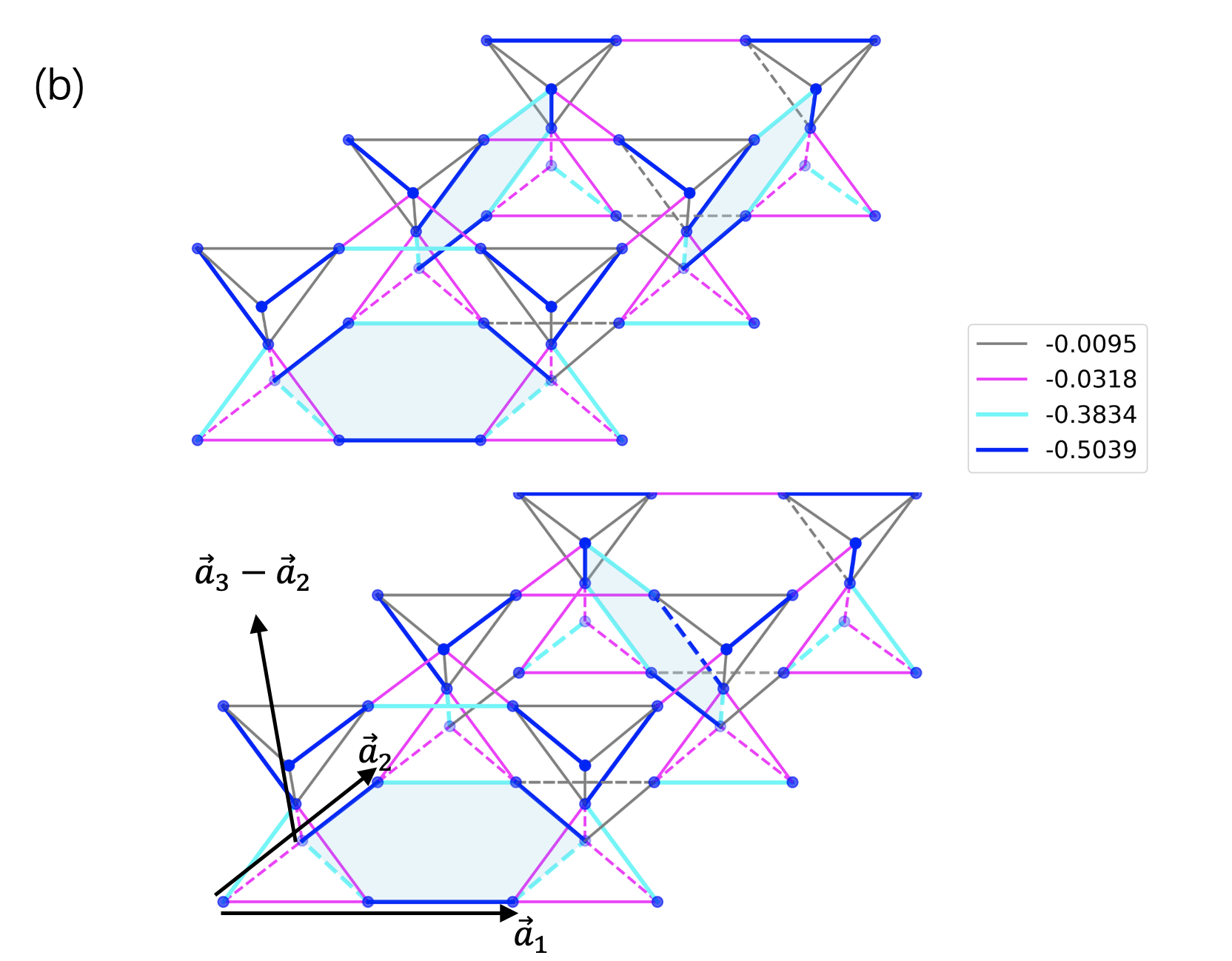}
\caption{(a)Illustration of a hard hexagon covering of the pyrochlore lattice with $2\vec{a}_{1}\times3\vec{a}_{2}\times2(\vec{a}_{3}-\vec{a}_{2})$ periodicity. A hard hexagon covering is composed of the $6$-spin loops directed in the directions of the four facets of an elementary tetrahedron in the pyrochlore lattice and is shown here in red, green, yellow and purple color respectively. (b) Illustration of the local spin correlation pattern in the fully optimized RVB state obtained by unrestricted optimization of the general RVB ansatz with the hard hexagon covering shown in (a) as the initial guess. For sake of clearance, we have illustrated two layers in the $\vec{a}_{3}-\vec{a}_{2}$ direction separately. Highlighted with light blue shaded area are resonating rings in the dressed hard hexagon VBC state. We note that while the NN spin correlation within the same resonating ring remains uniform in the variational state $|\Psi_{\alpha}\rangle=e^{-\alpha V}|\Psi_{0}\rangle$, it becomes dimerized in the fully optimized RVB state. The invariance of $|\Psi_{\alpha}\rangle$ under the combined operation of spatial inversion and translation or rotation is thus broken in the fully optimized state.}
\end{figure}   
 
From the figure we see that while the backbone of hard hexagon covering remains robust in the fully optimized state, the NN spin correlation within a hexagonal ring(as illustrated by the blue shaded areas in Fig.9b) becomes strongly dimerized. More specifically, the spin correlation on the stronger bond(-0.5039) is about 1.3 times larger than that on the weaker bond(-0.3834). A direct consequence of this dimerization is that $\langle \hat{\mathbf{S}}^{2}_{u} \rangle$ now becomes different for up and down tetrahedrons. More specifically, we have
\begin{equation}
\langle \hat{\mathbf{S}}^{2}_{u} \rangle \approx \left\{ 
\begin{aligned}
1.2119&, \ \ \ up \ tetrahedron\\ 
0.9085&, \ \ \  \ down \ tetrahedron\\
\end{aligned}\right.
\end{equation}
Thus the invariance of the dressed state $|\Psi_{\alpha}\rangle=e^{-\alpha V}|\Psi_{0}\rangle$ under the combined operation of spatial inversion ad translation or rotation is broken in the fully optimized state.    
     
Although we are still lack of a simple interpretation for such an astonishing degeneracy between the dressed hard hexagon VBC state and the maximally resonating VBC state, we note that they both feature 6-spin resonance on hexagonal rings. The key difference between the two states is as follows. While all sites are involved in such resonating rings and each site participates in only one resonating ring in the dressed hard hexagon VBC state, in the maximally resonating VBC state a site can participate in more than one resonating rings simultaneously, although not all sites are involved in such rings. More specifically, there is in total $\frac{N}{6}=\frac{2L^{3}}{3}$ resonating rings in the dressed hard hexagon VBC state. This is $\frac{2}{3}$ the number of resonating rings in the maximally resonating VBC state.  In addition, both types of VBC state feature strong disparity between up and down tetrahedrons after full optimization, although such a disparity is much stronger in the maximally resonating VBC state. With these close connections in mind, we can not help wondering if there is any continuous path in the Hilbert space that can connect these two seemingly very different VBC states smoothly. We leave such an interesting possibility to future study.

Finally, we note that the maximally resonating VBC state will be favored over the dressed hard hexagon VBC state when a tiny next-neighboring exchange coupling is turned on. More specifically, the spin correlation between next-nearest-neighboring sites in the fully optimized state are 
\begin{equation}
\sum_{j\in i_{nnn}}\langle \hat{\mathbf{s}}_{i}\cdot \hat{\mathbf{s}}_{j}\rangle \approx \left\{ 
\begin{aligned}
0.1583&, \ \ \  maximally \ resonating \ VBC\\ 
0.2022&, \ \ \  \ dressed \ hard \ hexagon \ VBC\\
\end{aligned}\right.
\end{equation}     
here $i_{nnn}$ denotes the next-nearest neighbor of $i$.

 \subsubsection{A comparison with the monopole flux state}
Finally, we compare the variational energy obtained here with that of the monopole flux state proposed in Ref.[\onlinecite{Han}] and Ref.[\onlinecite{Burnell}]. We note that on an equilateral pyrochlore cluster with periodic boundary condition, the mean field excitation spectrum of the monopole flux state is degenerate on the fermi level. To lift such a degeneracy and construct a well defined variational state, we impose anti-periodic boundary condition on two of the three basis vector directions. The size scaling of the variational energy for the monopole flux state is shown in Fig.8. From the plot we see the variational energy extrapolates nicely to a value of $-0.4574J/site$ in the thermodynamic limit. This energy is lower than the ideal isolated spin chain state mentioned before but is significantly higher than the maximally resonating VBC state found in this work. We note that the variational energy of the monopole flux state calculated here is consistent with the result reported in Ref.[\onlinecite{Han}] for the $L=4$ cluster, but is lower by a large margin than the result reported in Ref.[\onlinecite{Burnell}] for the $L=5$ cluster, which is $-0.4473J/site$. We think the discrepancy originates from the open shell problem of the mean field spectrum on the equilateral pyrochlore cluster.    

 \subsubsection{A summary of numerical results}
Collecting the numerical results presented in this section, we conclude that a symmetry breaking phase featuring a maximally resonating VBC pattern and $2\vec{a}_{1}\times2\vec{a}_{2}\times2\vec{a}_{3}$ periodicity emerges as a candidate ground state of the spin-$\frac{1}{2}$ PAFH. Such a unique VBC state is found to exhibit at least four levels of hierarchical structure in its symmetry breaking pattern. These hierarchical structures are driven by both the ice rule and spin resonance process at different orders. We also find that such a complicated VBC pattern can be qualitatively captured by an NN-RVB ansatz with the same $2\vec{a}_{1}\times2\vec{a}_{2}\times2\vec{a}_{3}$ periodicity. The extrapolated ground state energy of both the NN-RVB ansatz and the generalized RVB ansatz set new benchmarks on this complicated frustrated quantum magnet within the RVB framework. Intriguingly, we find that the maximally resonating VBC state is almost degenerate with a recently proposed dressed hard hexagon VBC state, when both are described in the RVB framework. We also find that after full optimization additional symmetry breaking will occur in the dressed hard hexagon VBC state and strong disparity in $\langle \hat{\mathbf{S}}^{2}_{u} \rangle$ will emerge between up and down tetrahedrons.

\begin{figure}
\includegraphics[width=8cm]{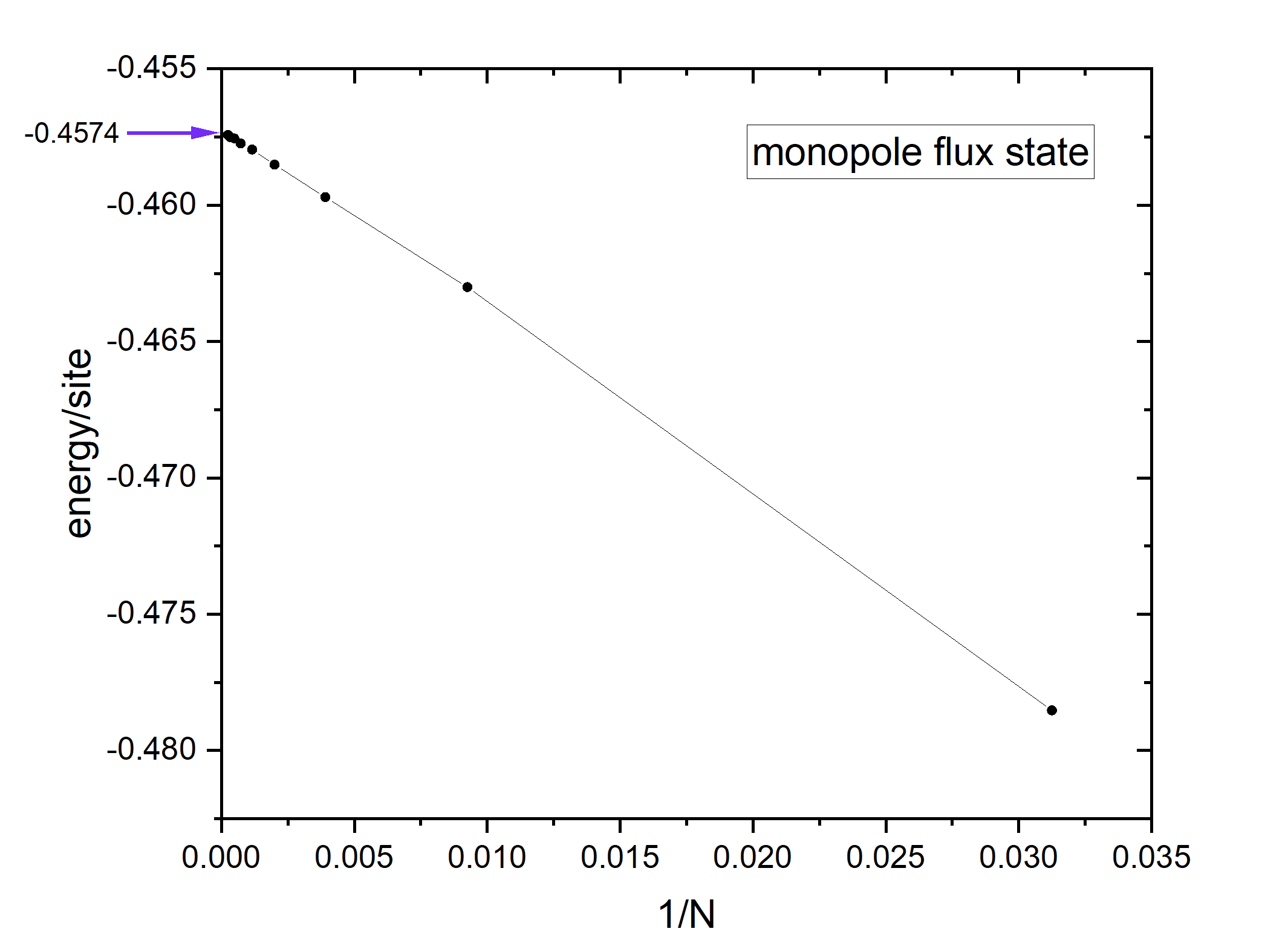}
\caption{Size scaling of the variational energy calculated from the mponopole flux state proposed in  Ref.[\onlinecite{Han}] and Ref.[\onlinecite{Burnell}]. The energy is calculated on an equilateral pyrochlore cluster with anti-periodic boundary condition on two of the three basis vector directions. The linear size of the cluster ranges from $L=2$ to $L=10$. $N=4L^{3}$ denotes the number of lattice site in the cluster.}
\end{figure}

\section{Conclusions and Discussions}
Elucidating the nature of the ground state of the spin-$\frac{1}{2}$ PAFH is a challenging problem as a result of the rapid scaling of the Hilbert space with the linear size of such a three dimensional system. After about three decades of intensive research the answer to this problem remains elusive. As the close analogy of the spin-$\frac{1}{2}$ KAFH, it has been conjectured by many researchers that the ground state of the spin-$\frac{1}{2}$ PAFH may also host a quantum spin liquid. However, evidence in support of a symmetry broken VBC ground state is also abundant in the literature. Reducing the bias of the method adopted in the study is the key to resolve the controversies about such a strongly frustrated system, for which close competition between different phases is naturally expected. In this work, we have performed a large scale unrestricted variational optimization for the ground state of the spin-$\frac{1}{2}$ PAFH within the resonating valence bond theory scheme. To reduce the variational bias, we have proposed the most general RVB ansatz that is consistent with the spin symmetry of the system. As the price we pay for the strong descriptive power of such a generalized RVB ansatz, we must optimize a huge number of variational parameters. More specifically, there is $N_{v}=4\times N^{2}$ variational parameters to be optimized on a finite pyrochlore cluster with $N$ sites. We have adopted the recently developed finite-depth BFGS algorithm to tackle such a challenging problem. The practice in this work lends further support to our earlier claim that this new algorithm can achieve good balance between numerical efficiency, numerical stability and storage demand. In particular, the new algorithm behaves very well even when we are dealing with system containing as many as $N=2048$ sites and wave function containing as may as $N_{v}=4\times N^{2}=16777216$ variational parameters.
    
We find a candidate ground state of the spin-$\frac{1}{2}$ PAFH that features a VBC pattern with $2\vec{a}_{1}\times 2\vec{a}_{2}\times 2\vec{a}_{3}$ periodicity. There are at least four levels of hierarchical structure in the symmetry breaking pattern of such a VBC state. At the first level of the hierarchy, all the down tetrahedrons(or all the up tetrahedrons) of the system are strongly spin dimerized. The spin dimerization is so strong that the spin correlation on the remaining bonds of the tetrahedron is almost suppressed to zero. As a result of such a spin dimerization, the valence bonds in the system fall roughly into three categories, namely the strong valence bonds in the down tetrahedron, the companying weak valence bonds in the down tetrahedron, and the moderate valence bonds in the up tetrahedron. Such a strong spin dimerization is driven by the need to minimize the total spin squared in the tetrahedrons. More specifically, we find that the stronger the spin dimerization in a tetrahedron, the smaller the expectation value of the total spin squared in that tetrahedron. At the second level of the hierarchy, the strong valence bond in $\frac{3}{4}$ of the down tetrahedrons form a maximally resonating pattern with $2\vec{a}_{1}\times 2\vec{a}_{2}\times 2\vec{a}_{3}$ periodicity. The enlarged unit cell thus contains $32$ sites. Such a complicated structure is driven by a $6$-spin resonance process around hexagonal rings composed of strong valence bonds in the $\frac{3}{4}$ down tetrahedrons and moderate valence bonds in up tetrahedrons. Above these two base hierarchical levels, these exist two higher level hierarchical structures in the VBC pattern which are related to the arrangement of the strong valence bonds in the remaining $\frac{1}{4}$ down tetrahedrons that do not participate in the $6$-spin resonance process and the arrangement of the moderate valence bonds in up tetrahedrons. These higher level hierarchical structures are driven by spin resonance process involving more spins and contribute very little to the ground state energy. Nevertheless, all these hierarchical structures can be resolved in fine details by our unrestricted optimization of the generalized RVB ansatz.

We find that an NN-RVB ansatz with $2\vec{a}_{1}\times 2\vec{a}_{2}\times 2\vec{a}_{3}$ periodicity can capture very well the qualitative feature of the maximally resonating VBC state. We find that the variational ground state energy obtained from both types of variational ansatz are very close to each other and extrapolate to $-0.4827J/site$ and $-0.4846J/site$ respectively in the thermodynamic limit. We note that these results are obtained on clusters containing as many as $N=2048$ sites and wave function containing as many as $N_{v}=16777216$ variational parameters. This is much larger than the scale that had ever been attempted in previous studies. As a byproduct, we have also falsified the monopole flux state proposed in Ref.[\onlinecite{Han}] and Ref.[\onlinecite{Burnell}]. According to our calculation, the variational energy of the monopole flux state extrapolates to $-0.4574J/site$ in the thermodynamic limit. This is much higher than the maximally resonating VBC state found in this work.

We note that in the semiclassical limit of $S\rightarrow \infty$, the PAFH is more strongly frustrated than the KAFH. This can be seen by counting the degree of freedom in the ground state manifold of both model as dictated by the ice rule of ${\mathbf{S}}_{u}=0$, which is extensive for PAFH but sub-extensive for the KAFH. However, while the ground state of the spin-$\frac{1}{2}$ KAFH is generally believed to be a quantum spin liquid, the ground state of the spin-$\frac{1}{2}$ PAFH is found here to feature a symmetry breaking VBC pattern. The frustration level of a model in the semiclassical limit is thus not a good indicator for the potential of the corresponding quantum model to realize a quantum spin liquid ground state. In fact, while the ice rule can still be satisfied locally for the spin-$\frac{1}{2}$ PAFH, it can never be satisfied for the spin-$\frac{1}{2}$ KAFH. We think this is the key difference between these two strongly frustrated quantum antiferromagnets. Indeed, our results indicate that the strong spin dimerization in the ground state of the spin-$\frac{1}{2}$ PAFH is just driven by the need to optimize such an ice rule condition. In this sense the spin-$\frac{1}{2}$ PAFH is less frustrated than the spin-$\frac{1}{2}$ KAFH. 

Through large scale unrestricted variational optimization, we also find an intriguing degeneracy between the maximally resonating VBC state and a recently proposed dressed hard hexagon VBC state, although they have very different structures. Currently it is not clear what is the origin for such a degeneracy. However, we note that both VBC states feature $6$-spin resonance on hexagonal rings of the pyrochlore lattice, although there is $\frac{3}{2}$ times more such resonating rings in the maximally resonating VBC state. At the same time, we find that further symmetry breaking will occur in the dressed hard hexagon VBC state under unrestricted optimization and there is strong disparity in $\langle \hat{\mathbf{S}}^{2}_{u} \rangle$ for up and down tetrahedrons, as we have observed in the maximally resonating VBC state. Thus, no matter which state will eventually be selected as the ground state of the spin-$\frac{1}{2}$ PAFH, we expect that $\langle \hat{\mathbf{S}}^{2}_{u} \rangle$ to be different for up and down tetrahedron. Such a tendency will resonate coherently with the tendency toward breathing distortion commonly observed in pyrochlore materials.      
Finally, we find that the maximally resonating VBC state can be favored over the dressed hard hexagon VBC state when a tiny next-nearest neighboring exchange coupling is turned on.

The results reported here may have direct relevance to the pyrochlore spin liquid candidate material R$_{2}$Mo$_{2}$O$_{5}$N$_{2}$, in particular the recently synthesized Lu$_{2}$Mo$_{2}$O$_{5}$N$_{2}$. Here the spin-$\frac{1}{2}$ Mo$^{5+}$ ions occupying the pyrochlore lattice site is believed to have a dominant antiferromagnetic Heisenberg exchange coupling between nearest neighboring sites\cite{Clark,Iqbal3}. Recent measurement on this material find that the system remains paramagnetic at a temperature scale much smaller than that set by the exchange coupling. Our results may also have relevance to the pseudospin $S=\frac{1}{2}$ Yb-based compound Ba$_{3}$Yb$_{2}$Zn$_{5}$O$_{11}$, in which the spin-$\frac{1}{2}$ Yb$^{3+}$ ions occupy the site of a breathing pyrochlore lattice\cite{breathing1,breathing2}. The strength of the Heisenberg exchange coupling within the up and the down tetrahedrons are thus by construction different. The strong tendency toward spin dimerization in the ground state of the spin-$\frac{1}{2}$ PAFH found in this work matches perfectly with such a breathing structure in the lattice. It is thus reasonable to expect that the hierarchical structure in the VBC pattern found in this work may be realized in such a breathing pyrochlore antiferromagnet. More generally, the pyrochlore antiferromagnet form a large material family. Generalized PAFH with exchange anisotropy, longer-ranged exchange coupling and higher spin can be easily envisaged. Although the results of this work may not be applicable to these systems directly, there is no conceptual difficulty to generalize the approach adopted in this work to study these models. We leave these open problems to future study.

\begin{acknowledgments}
We acknowledge the support from the National Natural Science Foundation of China(Grant No.12274457). We are grateful to Alaric Sanders and Robin Schaefer for correspondences.
\end{acknowledgments}

\end{document}